%% file: arXiv_version.tex
\documentclass[final,twocolumn,5p,times,10pt,authoryear]{elsarticle}
\usepackage{lineno}
\modulolinenumbers[5]

\journal{Automatica}

\usepackage[T1]{fontenc}
\usepackage[utf8]{inputenc}
\usepackage{bm}
\usepackage{hhline}
\usepackage{csquotes}
\usepackage{xcolor}
\usepackage{algorithm}
\usepackage{dsfont}
\usepackage{cite}
\usepackage{amsmath,amssymb,amsfonts}
\usepackage{algorithmic}
\usepackage{textcomp}
\usepackage{graphicx}
\usepackage[math]{cellspace}
\usepackage{diagbox} 
\usepackage{pict2e}
\usepackage{subeqnarray}
\usepackage{import}
\usepackage{color,epstopdf,amssymb,multirow,url,soul}
\usepackage{hyperref}
\usepackage{mathtools}
\usepackage{soul}

\setstcolor{magenta}
\newcommand{\GR}[1]{\textcolor{black}{#1}}
\newcommand{\VL}[1]{\textcolor{black}{#1}}

\newcommand{\be}{\begin{equation}}
\newcommand{\ee}{\end{equation}}

\newcommand{\matriz}[1]{
\begin{bmatrix}
  #1
\end{bmatrix}
}

\newcommand\numberthis{\addtocounter{equation}{1}\tag{\theequation}}

 \newtheorem{pro}{\noindent Problem}
\newtheorem{proof}{\noindent Proof}

\newtheorem{theorem}{\bf{Theorem}}
\newtheorem{lemma}{\bf{Lemma}}
\newtheorem{proposition}{\bf{Proposition}}
\newtheorem{definition}{\bf{Definition}}
\newtheorem{remark}{\bf{Remark}}
\newtheorem{assumption}{\bf Assumption}

\biboptions{authoryear}

\allowdisplaybreaks 

\begin{document}

\begin{frontmatter}

\title{Stabilizing NMPC Approaches for Underactuated Mechanical Systems on the $\mathsf{SE(3)}$ Manifold\tnoteref{mytitlenote}}

\tnotetext[mytitlenote]{\copyright 2025. This manuscript version is made available under the CC-BY-NC-ND 4.0 license \href{https://creativecommons.org/licenses/by-nc-nd/4.0/}{https://creativecommons.org/licenses/by-nc-nd/4.0/}. This paper was not presented at any IFAC meeting. This work was partially supported by the Brazilian agencies CAPES through the Academic Excellence Program (PROEX), CNPq under the grants 317058/2023-1, 422143/2023-5, 311891/2022-5 and FAPEMIG under the grant APQ-02100-22, and Project Stic-Amsud/CAPES NetConHybSDP (20/2022).}

\author[cefet]{Jean C. Pereira\corref{mycorrespondingauthor}}
\cortext[mycorrespondingauthor]{Corresponding author: \texttt{jean@cefetmg.br} }   
\author[cefet]{Valter J. S. Leite}
\author[Raffo]{Guilherme V. Raffo}
\address[cefet]{Department of Mechatronics Engineering, {\em Campus} Divin\'{o}polis --- CEFET--MG, Divin\'{o}polis, MG, 35503-822, Brazil}                    

\address[Raffo]{Department of Electronics Engineering, Federal University of Minas Gerais, Belo Horizonte, MG 31270-901, Brazil}        

\begin{abstract}
\GR{This paper addresses the motion control problem for underactuated mechanical systems with full attitude control and one translational force input to manage the six degrees of freedom involved in the three-dimensional Euclidean space. These systems are often classified as second-order nonholonomic due to their completely nonintegrable acceleration constraints. To tackle this complex control problem, we propose two nonlinear model predictive control (NMPC) schemes that ensure closed-loop stability and recursive feasibility without terminal conditions. The system dynamics are modeled on the $\mathsf{SE(3)}$ manifold for a globally and unique description of rigid body configurations. One NMPC scheme also aims to reduce mission time as an economic criterion. The controllers' effectiveness is validated through \VL{numerical experiments on} a quadrotor UAV.}
\end{abstract}


\end{frontmatter}

\section{Introduction}
Unmanned vehicles have \GR{gained significant attention recently due to their wide range of applications, including security, agriculture, infrastructure inspection, exploration, wildlife conservation, package delivery, and remote sensing 
\citep{He2022,Penicka2022,CENERINI2023,Hamrah2018}.
Depending on their environment, these vehicles are categorized as Autonomous Ground Vehicles (AGVs), Autonomous Underwater Vehicles (AUVs), and Unmanned Aerial Vehicles (UAVs) \citep{JLiu2017,Heshmati-Alamdari2021,HUA2013}. Typically, they are designed with fewer actuators than degrees of freedom (DOF) to reduce weight and cost, making them underactuated systems \citep{He2022}.}

\GR{This work investigates the motion control of underactuated mechanical systems globally represented in the $\mathsf{SE(3)}$ manifold, which features} full attitude control \GR{with a single} translational input, \GR{using} four independent control inputs to manage the six DOF in the three-dimensional Euclidean space. These systems are often classified as second-order nonholonomic due to their nonintegrable acceleration constraints. \VL{
The Special Euclidean group $\mathsf{SE(3)}$ \GR{is defined as an ordered pair $(R,\xi) \in \mathsf{SE(3)}$, where $R \in \mathsf{SO(3)}$\footnote{\GR{The notation $\mathsf{SO}$ abbreviates \textit{special orthogonal}. Generally, the space of rotation matrices in $\mathbb{R}^{n\times n}$ is given by $\mathsf{SO}(n)\coloneqq\{R\in \mathbb{R}^{n\times n}:RR^\top=I,\textrm{det} R=+1\}$ \citep{Murray94}.}}
is a rotation matrix indicating the body's attitude, and $\xi \in \mathbb{R}^3$ denotes the position of a point, typically the origin of the body-fixed frame, in the inertial frame \citep{Lee2018}.} Minimal representations, \GR{like Euler-angle and angle-axis parametrizations, face singularities and are only locally effective \citep{Lee2015}. In contrast, unit quaternions are singularity-free \citep{CASAU2015} and  provide a double cover of the attitude space, leading to potential inconsistencies such as} the unwinding phenomenon \citep{Mayhew2011}. \GR{Thus, using rotation matrices and the $\mathsf{SE(3)}$ manifold is advantageous for controlling underactuated mechanical systems, particularly for aggressive motions, as this representation is unique and independent of local coordinates \citep{Chaturvedi2011,He2022}, effectively capturing the coupling between rotational and translational dynamics \citep{Nikhilraj2019}.}} 
\VL{A $\mathsf{SE(3)}$ manifold} model applies to a wide range of rotary-wing aircraft, including helicopters \citep{Frazzoli2000}, multi-propeller helicopters 
\citep{HUA2013}, and ducted fans \citep{Pflimlin2007}, as well as satellites \citep{YOSHIMURA2016}. 
\GR{Accordingly, since the resultant force vector typically aligns with a body-fixed direction, the control of attitude and translational dynamics must be integrated simultaneously} \citep{Hamrah2018}.

\GR{Stabilizing this class of underactuated mechanical systems is challenging, as nonholonomic systems with rotational DOF} cannot be globally asymptotically stabilized at a constant equilibrium via continuous feedback \citep{Brockett83,Bhat2000}. \GR{To address this, several solutions have been proposed, including discontinuous control laws on $\mathsf{SE(3)}$ \citep{Fjellstad94,Egeland96}, time-varying quaternion continuous feedback \citep{Pettersen99}, and hybrid quaternion feedback strategies 
\citep{CASAU2015}.} However, these \GR{approaches} disregard control input constraints, which are crucial in practical applications \citep{Huiping2017}.


\GR{To address the challenges of controlling underactuated vehicles,} model predictive control (MPC) schemes have been \GR{employed due to their capability to handle constrained multiple-input multiple-output nonlinear systems efficiently \citep{raffo2010,Kunz2013,Alexis2014,Huiping2017,Eskandarpour2019,Liang2021, CENERINI2023, Iuru2023}. However, many studies} have used local attitude parametrization, like Euler angles, and separated the problem into attitude and position control layers \citep{raffo2010,Alexis2014,Eskandarpour2019}, which can be limiting for tasks involving simultaneous position and attitude references. Addressing \VL{such limitation}, \citet{Murilo2019} have introduced a unified NMPC framework using Euler angles for quadrotor UAVs, but it can also suffer from singularities and be limited to local effectiveness. \GR{Similarly, \citet{Wehbeh2022} have proposed a unified trajectory tracking scheme using time-dependent linearization on the Lie algebra of $\mathsf{SO(3)}$ to avoid singularities. Despite achieving applicable results, the control strategy's feasibility and closed-loop stability} have not been demonstrated.

For simplified and linearized models, such as in \citet{Eskandarpour2019}, feasibility and closed-loop stability analyses of MPC are typically provided. However, these analyses are limited for NMPC strategies considering a complete nonlinear model of underactuated vehicles without decoupling rotational and translational motions \citep{Liang2021}. \GR{Classical MPC with terminal constraints usually requires a locally continuous time-invariant state feedback controller for system stability \citep{Mayne2000}, but Brockett's theorem \citep{Brockett83} states that there is no such feedback controller for nonholonomic systems. To address this, \citet{Huiping2017} have proposed an NMPC scheme using a transformed model with homogeneity properties for underactuated marine vehicles, although it requires \VL{changing the control law back.} 
In \citet{Liang2021}, an NMPC with terminal constraints developed using Lyapunov's stability theory and backstepping has been presented for marine vehicles in two-dimensional Euclidean space, introducing an auxiliary time-varying tracking controller.} Another strategy to overcome the stability requirement is focusing on NMPC frameworks without terminal conditions.


\GR{MPC schemes without stabilizing constraints are easier to design, require less computational effort, and are commonly used in industrial applications compared to those with stabilizing constraints} \citep{MEHREZ2020}. \GR{Furthermore, they provide} larger stability regions for a fixed prediction horizon \citep{Grune2017}. In this type of MPC, closed-loop stability is typically ensured by \VL{indirect techniques} using bounds on the MPC value function to determine a stabilizing prediction horizon length \citep{Tuna2006,Grune2010}, or \GR{through recent advancements in economic MPC (EMPC) that leverage} conditions based on dissipativity and controllability properties \citep{GRUNE2014}. 
%
%
\GR{In EMPC,} the controller's primary goal is not to stabilize a precomputed trajectory or steady state but \GR{to optimize performance based on economic cost criteria, such as minimizing energy or time} \citep{Grune2020}. 


\GR{This work addresses the motion control problem of a class of underactuated mechanical systems in the non-Euclidean space $\mathsf{SE(3)}$. 
The closed-loop stability analysis relies on dissipativity properties typically associated with EMPC frameworks. Therefore, we propose two NMPC schemes without terminal conditions to tackle the stabilization problem on  the $\mathsf{SE(3)}$ manifold. The first scheme regulates the vehicle's pose to achieve a desired equilibrium point, while the second incorporates mission time as an economic criterion, retaining the features of the first controller. This approach utilizes an exponential weight in the stage cost to indirectly achieve time-optimality, adapting the fixed horizon strategy proposed by \citet{Verschueren2017}.}

\GR{The contributions of the paper are threefold: i) the development of NMPC schemes for the stabilization of a class of underactuated mechanical systems with second-order nonholonomic constraints globally represented on the $\mathsf{SE(3)}$ manifold, capable of addressing constraints on inputs and states; ii) guarantees of closed-loop stability and recursive feasibility for both schemes,  \textcolor{black}{derived from a dissipativity-based analysis that extends established principles (\citet{Tuna2006}; \citet{Grune2010}; \citet{GRUNE2014}; \citet{Grune2017}) to second-order nonholonomic systems on the SE(3) manifold, where unique challenges arise from state-space representation and constraints}; and iii) the introduction of an NMPC scheme that considers mission time reduction as an economic criterion\textcolor{black}{, which is particularly significant for practical applications like UAV navigation and robot motion, where operational efficiency is critical.}}

\subsection*{\VL{Notations}}\label{subsec:not}

$\mathbb{R}$, $\mathbb{R}^n$, $\mathbb{R}^{m \times n}$, and $\mathbb{R}^+_0$ denote the spaces of real numbers, vectors with $n$ real entries, matrices of dimension $m \times n$ with real entries, and the non-negative real numbers, respectively. The identity matrix is represented by $I_n$, while zero matrices are denoted by $\textbf{0}_{m \times n}$ and quadratic zero matrices by $\textbf{0}_n$. 
The transpose of a matrix $M$ is $M^\top$, and for a square matrix, the transpose of its inverse is $M^{-\top}$. 
The set of integers from $0$ to $i$ is denoted by $\mathbb{I}_{0:i}$. A continuous function $\alpha : \mathbb{R}^+_0 \rightarrow \mathbb{R}^+_0$ is a class $\mathcal{K}$ function if it is zero at zero and strictly monotonically increasing; it is a class $\mathcal{K}_\infty$ function if unbounded. A continuous function $\beta : \mathbb{R}^+_0 \times \mathbb{R}^+_0 \rightarrow \mathbb{R}^+_0$ is a $\mathcal{KL}$-function if $\beta(\cdot, t) \in \mathcal{K}$ for all $t \in \mathbb{R}^+_0$ and $\beta(r,\cdot)$ strictly monotonically decays to zero for each $r>0$. The vector space of skew matrices is denoted by $\mathfrak{so}(\cdot)$, and $\mathcal{S}(\cdot)$ represents the skew-symmetric matrix operator. Time dependence of a variable $v$ is denoted by $v(t)$ or $v(k)$ for clarity.

\section{Problem Formulation\label{sec:prb}}

Consider a finite-dimensional nonlinear system of the form
\small
\begin{gather}
    \dot{x}(t)=g(x(t),u(t)), \label{fg_ps}
\end{gather}
\normalsize
for some state-transition map $g: X \times U \rightarrow X$ \textcolor{black}{continuously differentiable}, with $X$ and $U$ being the state and control normed spaces, respectively. Additionally, $x \in \mathbb{X} \subseteq X$ represents the states and $u \in \mathbb{U} \subseteq U$ the inputs, with $\mathbb{X}$ and $\mathbb{U}$ designating, respectively, the set of admissible states and inputs.
\color{black}
The discrete-time system $x^+ = g_d(x,u)$ is obtained from \eqref{fg_ps} under zero-order-hold control with sampling period $T_s$:
\small
\begin{gather}\label{fg_ps_dis}
    x(j+1)=g_d(x(j),u(j)) = \phi(t+T_s;t,x,u),
\end{gather}
\normalsize
where $\phi(\tau;t,x,u)$ denotes the solution at time $\tau$ of \eqref{fg_ps} with initial condition $x(t) = x$ and constant control $u$.
\color{black}
The discrete-time optimal control problem (OCP) to be solved at each time instant $k$, with $t=kT_s$, is posed as
\small
\begin{gather}
\label{ocp_gen}
\begin{matrix}
	\displaystyle
	\min_{x,u}\sum_{j=0}^{N_p}\ell(x(k+j),u(k+j))\\  
	\begin{array}{rll}
	\textrm{s.t.} & x(j+1)=g_d(x(j),u(j)), & x_0=x(k),\\
	 &x \in \mathbb{X}, & \forall j\in \mathbb{N}_p,\\
	 & u \in\mathbb{U}, &  \forall j\in \mathbb{N}_p,
	\end{array} 
	\end{matrix}
\end{gather}
\normalsize
where $\ell: X \times U \rightarrow \mathbb{R}^+_0$ is the \textcolor{black}{discrete-time} stage cost, \textcolor{black}{defined as the temporal average of a continuous-time cost $\ell_c(x,u)$:
\small
\begin{gather}\label{cost_def}
    \ell(x,u) = \frac{1}{T_s} \int_t^{t+T_s} \ell_c(\phi(\tau;t,x,u), u) d\tau.
\end{gather}
\normalsize 
Moreover,} $x_0$ is the current state, and $\mathbb{N}_p$ represents the non-negative integers bounded by the prediction horizon $N_p$. The optimization variables are given by the sequences $x\coloneqq (x(0),x(1),\dots,x(N_p))$ and $u\coloneqq(u(0),u(1),\dots,u(N_p-1)$.

The following MPC iteration defines the feedback law at each time instant $k$: i) Measure the current state $x_0$; ii) Solve the OCP \eqref{ocp_gen} to obtain the optimal control sequence $u^\star$; iii) Apply the first control action $u^\star(0)$ until the next time instant.

\begin{definition}[adapted from \citet{GRUNE2014}]
A pair $(x^e,u^e) \in \mathbb{X} \times \mathbb{U}$ is called equilibrium for the control system \eqref{fg_ps_dis} if it satisfies the condition $g_d(x^e,u^e)=x^e$. An equilibrium is optimal if it solves the optimization problem
\small
\begin{gather}
\min_{x\in \mathbb{X},u\in \mathbb{U}}\ell(x,u)~~~~	\textrm{s.t. } g_d(x,u)-x=0 \ .
\end{gather}
\normalsize
\end{definition}
Therefore, for the class of underactuated mechanical systems with second-order nonholonomic constraints
, with the state-transition map $g$ defined by \eqref{ms1}-\eqref{ms4}, the following problems are addressed:
\begin{pro}\label{prob1}
    Steer the underactuated mechanical system to a desired set-point $x^e$, which is an optimal equilibrium for the control system.
\end{pro}
\begin{pro}\label{prob2}
    Solve Problem \ref{prob1} regarding the mission time as an economic criterion.
\end{pro}

\section{Preliminaries}

\subsection{Dynamics of a class of underactuated mechanical systems on $\mathsf{SE(3)}$}\label{subsec:dyn}
The system is modeled as an underactuated rigid body \GR{with a frame $\mathcal{B}=\{B_1,B_2,B_3\}$ attached} to it. The inertial frame $\mathcal{I}=\{I_1,I_2,I_3\}$ is fixed with respect to the Earth. The configuration manifold is represented by the ordered pair $(R^{\mathcal{I}}_{\mathcal{B}},\xi^{\mathcal{I}}) \in$ $\mathsf{SE(3)}$, where the rotation matrix $R^{\mathcal{I}}_{\mathcal{B}} \in$ $\mathsf{SO(3)}$ describes the rigid body attitude, and $\xi^{\mathcal{I}} \in \mathbb{R}^3$ denotes the center of mass position of the vehicle. The state space of the system is $\mathsf{TSE(3)}$, \GR{the tangent bundle of $\mathsf{SE(3)}$, allowing the motion dynamics to be represented through the states} $x \coloneqq (R^{\mathcal{I}}_{\mathcal{B}},\xi^{\mathcal{I}},\omega^{\mathcal{B}},v^{\mathcal{I}}) \in$ $\mathsf{TSE(3)}$. The continuous-time equations of motion are given by \citep{Kobilarov2013}
\small
\begin{align}
\dot{\xi}^{\mathcal{I}} &= v^{\mathcal{I}},\label{ms1} \\  
m\dot{v}^{\mathcal{I}} &= f^{\mathcal{I}}= R^{\mathcal{I}}_{\mathcal{B}}e^{\mathcal{B}}T +f_e^{\mathcal{I}}(\xi^{\mathcal{I}},v^{\mathcal{I}}),\label{ms2} \\  
\dot{R}^{\mathcal{I}}_{\mathcal{B}}&=R^{\mathcal{I}}_{\mathcal{B}}\mathcal{S}(\omega^{\mathcal{B}}),\label{ms3}\\ 
\textcolor{black}{J^{\mathcal{B}}}\dot{\omega}^{\mathcal{B}}&=-\omega^{\mathcal{B}}\times J^{\mathcal{B}}\omega^{\mathcal{B}}+\tau^{\mathcal{B}}+\tau_{e}^{\mathcal{B}}(x), \label{ms4}
\end{align}
\normalsize
where $T\in \mathbb{R}$ is a control force applied in a body-fixed direction defined by the unit vector $e^{\mathcal{B}}\in\mathbb{R}^3$, $\tau^{\mathcal{B}}\in\mathbb{R}^3$ is the control torque expressed in 
$\mathcal{B}$, both applied at the motion system center of mass, $v^{\mathcal{I}} \in \mathbb{R}^3$ is the center of mass linear velocity expressed in $\mathcal{I}$, $\omega^{\mathcal{B}} \in \mathbb{R}^3$ is the angular velocity of the body with respect to $\mathcal{I}$, expressed in $\mathcal{B}$, $\mathcal{S}(\omega^{\mathcal{B}}) \in$ $\mathfrak{so}(3)$, $m$ is the total mass of the system, and \textcolor{black}{$J^{\mathcal{B}}$} is the inertia tensor. The vehicle is subjected to known external forces and torques given by the functions $f_e^{\mathcal{I}} : \mathbb{R}^3 \times \mathbb{R}^3  \rightarrow  \mathbb{R}^3$ and $\tau_{e}^{\mathcal{B}}: \textrm{T$\mathsf{SE(3)}$}\rightarrow \mathbb{R}^3$, respectively.
\subsection{Attitude error function on $\mathsf{SO(3)}$}

Although a system modeled by a rotation matrix $R \in$ $\mathsf{SO(3)}$ has no representational singularities, selecting a positive definite error function to measure the difference between configurations is \GR{crucial for} the control design. \GR{Consider the $i$-th configuration error function {$\Psi_i(R,R_d)=\frac{1}{2}\|Re_i-R_de_i\|^2=1-Re_i\cdot R_de_i$
},  
where $e_i \in \mathsf{S}^2 \coloneqq \{q\in\mathbb{R}^3:\|q\|^2=1\}$ is a unit-vector starting from the mass center of the rigid body~\citep{Lee2015}. Let $e_1$ and $e_2$ be orthogonal vectors.} 
For positive constants $k_i>0$, the complete configuration error function is given by
\small
\begin{gather}
\Psi(R,R_d)=k_1\Psi_1(R)+k_2\Psi_2(R).
\label{psiR}
\end{gather}
\normalsize
It can be noticed that $\Psi_i(R,R_d)$ represents the geometric projection between $Re_i$ and $R_de_i$, and $\Psi_i(R,R_d)$ is positive definite about $Re_i= R_de_i$. 
Therefore, the complete configuration error $\Psi(R,R_d)$, given by \eqref{psiR}, is positive definite about $R=R_d$, that is, {$\Psi(R,R_d)>0$ for all $R\neq R_d$,} and $\Psi(R,R_d)=0$ if only if $R=R_d$.

\subsection{System discretization}
The Euler method is used to discretize the dynamic equations \eqref{ms1} and \eqref{ms2}, with the predicted time $t_p = jT_s$, resulting in
\small
\begin{align}
   \xi^{\mathcal{I}}(j+1) & = \xi^{\mathcal{I}}(j) + T_s v^{\mathcal{I}}(j)\label{discrete01},\\ 
   v^{\mathcal{I}}(j+1) & = v^{\mathcal{I}}(j) + T_s \left(\frac{R^{\mathcal{I}}_{\mathcal{B}}(j)e^{\mathcal{B}}T(j) +f_e^{\mathcal{I}}(j)}{m}\right) \label{discrete02}.
\end{align}
\normalsize
The traditional Euler method does not ensure that the predicted $R^{\mathcal{I}}_{\mathcal{B}}(j+1)$ remains on $\mathsf{SO(3)}$. Consequently, specific geometric integration methods \GR{have been developed for numerically integrating} \eqref{ms3} and \eqref{ms4} \citep{giuliatese}. \GR{These methods primarily rely} on the geometric relationship between $\mathsf{SO(3)}$ and $\mathfrak{so}(3)$, with $\mathfrak{so}(3)$ representing the tangent space of $\mathsf{SO(3)}$ at the identity $I_3 \in \mathsf{SO(3)}$. Since $\mathfrak{so}(3)$ is the Lie algebra of the Lie Group $\mathsf{SO(3)}$, the map $\mathfrak{so}(3)\rightarrow\mathsf{SO(3)}$ corresponds to the exponential matrix map. Thus, the solution to the differential equation \eqref{ms3} is given by \GR{the exponential map $R^{\mathcal{I}}_{\mathcal{B}}(t)=e^{t\mathcal{S}(\omega^{\mathcal{B}}(t))}=I_3+t\mathcal{S}(\omega^{\mathcal{B}}(t))+\frac{(t\mathcal{S}(\omega^{\mathcal{B}}(t)))^2}{2!}+\dots$, 
which can be discretized as $R^{\mathcal{I}}_{\mathcal{B}}(j)=R^{\mathcal{I}}_{\mathcal{B}}(jT_s)=e^{jT_s\mathcal{S}(\omega^{\mathcal{B}}(jT_s))}$.} 
Thus, given the current configuration at the $j$-th sample time ($R^{\mathcal{I}}_{\mathcal{B}}$($j$), $\omega^{\mathcal{B}}$($j$)), the next configuration ($R^{\mathcal{I}}_{\mathcal{B}}$($j$+1), $\omega^{\mathcal{B}}$($j$+1)) is defined by the following iteration rule \citep{simo88}:
\small
\begin{align}
    \omega^{\mathcal{B}}(\bar{j})&=\omega^{\mathcal{B}}(j)+\frac{T_s}{2}\textcolor{black}{(J^{\mathcal{B}}})^{-1}\alpha(j),\label{eq:Rdyamics03}\\
    R^{\mathcal{I}}_{\mathcal{B}} (j+1)&=R^{\mathcal{I}}_{\mathcal{B}}(j)\mbox{cay}(T_s\omega^{\mathcal{B}}(\bar{j})), \label{ln2}\\
    \omega^{\mathcal{B}}(j+1)&=\omega^{\mathcal{B}}(\bar{j})+\frac{T_s}{2}\textcolor{black}{(J^{\mathcal{B}}})^{-1}\alpha(j) \label{eq:Rdyamics04},
\end{align}
\normalsize
where $\alpha(j)=\VL{J^{\mathcal{B}}}\omega^{\mathcal{B}}(j)\times \omega^{\mathcal{B}}(j)+\tau^{\mathcal{B}}(j) + \tau_{e}^{\mathcal{B}}(j)$, and cay$(\cdot)$ is the Cayley map, which is the second-order approximation of the matrix exponential function defined by \GR{$\textrm{cay}(\nu) \coloneqq (I_3+\frac{1}{2}\mathcal{S}(\nu))(I_3-\frac{1}{2}\mathcal{S}(\nu))^{-1} \approx \textrm{expm}(\mathcal{S}(\nu))$.}

\subsection{MPC Stability Based on Strict Dissipativity}\label{section:stabity_background}
The main dissipativity-based stability \textcolor{black}{results, definitions, and assumptions} for MPC schemes without terminal conditions presented in \citet{GRUNE2014} are revisited in this section (see also \citet{Grune2017} for further details).
\begin{definition}
Consider $x^e\in \mathbb{X}$ to be an equilibrium for the closed-loop system. The equilibrium is said to be practically asymptotically stable w.r.t. $\rho \geq 0$ on a set $S \subseteq \mathbb{X}$ with $x^e \in S$ if there exists $\beta \in \mathcal{KL}$ such that
\small
\begin{gather}\label{pract}
    \|x-x^e\|\leq\max\{\beta(\|x_0-x^e\|,\textcolor{black}{k}),\rho\}
\end{gather}
\normalsize
holds for all $x_0 \in S$ and all $\textcolor{black}{k} \in \mathbb{N}$. The equilibrium is globally practically asymptotically stable w.r.t. $\rho\geq 0$ if \eqref{pract} holds on $S = \mathbb{X}$.
\end{definition}

\begin{assumption}\label{ass:dissi}
The OCP \eqref{ocp_gen} is strictly dissipative with respect to the equilibrium $(x^e,u^e) \in \mathbb{X} \times \mathbb{U}$, that is, there is a non-negative storage function $\lambda:\mathbb{X}\rightarrow\mathbb{R}^+_0$, and a function $\alpha_\ell \in \mathcal{K}_\infty$ such that
\small
\begin{gather}\label{dissipa:discrete}
    \lambda(g_d(x,u))-\lambda(x) \leq \ell(x,u)-\ell(x^e,u^e)-\alpha_\ell(\|x-x^e\|),
\end{gather}
\normalsize
holds for all $\textcolor{black}{x} \in \mathbb{X}$, $u\in\mathbb{U}$, $j \in \mathbb{N}$. 
\end{assumption}
\color{black}
\begin{assumption}\label{ass:regularity}
The state and control constraint sets $\mathbb{X}$ and $\mathbb{U}$ are compact. The stage cost $\ell$ is bounded on $\mathbb{X} \times \mathbb{U}$. The storage function $\lambda$ and $\alpha_\ell$ are Lipschitz continuous on a ball $\mathcal{B}_\epsilon(x^e) \coloneqq \{x \in \mathbb{X} \mid \|x-x^e\|<\epsilon\}$ for some $\epsilon>0$.
\end{assumption}

\begin{assumption}\label{ass:local_control}[Adapted from \citet[Assumption 3.3]{GRUNE2014}] The system \eqref{fg_ps_dis} is locally controllable on $\mathcal{B}_{\varepsilon}(x^{e})$. Specifically, there exist $\epsilon > 0$, $N' \in \mathbb{N}$, and $C > 0$ such that for all $x_{0} \in \mathcal{B}_{\varepsilon}(x^{e})$, there exist $u_{1} \in \mathbb{U}^{N'}(x_{0})$ and $u_{2} \in \mathbb{U}^{N'}(x^{e})$ satisfying $x_{u_{1}}(N', x_{0}) = x^{e}$ and $x_{u_{2}}(N', x^{e}) = x_{0}$, where $x_{u_{i}}(N', x_{0})$ denotes the state trajectory of \eqref{fg_ps_dis} resulting from a control sequence $u_{i}\in\mathbb{U}^{N^{\prime}}(x_{0})$ starting at $x_{0}\in\mathbb{X}$, with $\mathbb{U}^{N'}(x_{0})$ being the set of all admissible control sequences which holds both $u \in \mathbb{U}$ and $x_{u}(j, x_{0}) \in \mathbb{X}$ for all $j \in \mathbb{I}_{0:N'}$. Furthermore, the following bound holds for all $j \in \mathbb{I}_{0:N'-1}$: $\max\left\{\|x_{u_{1}}(j, x_{0}) - x^{e}\|, \|x_{u_{2}}(j, x^{e}) - x^{e}\|, \|u_{1}(j) - u^{e}\|, \|u_{2}(j) - u^{e}\|\right\} \\ \leq C\|x_{0} - x^{e}\|$. 
\end{assumption}
\begin{assumption}\label{ass:finite_control}[Adapted from \citet[Assumption 3.4]{GRUNE2014}] The system \eqref{fg_ps_dis} is finite-time controllable into $\mathcal{B}_{\varepsilon}(x^{e})$. Specifically, for the $\epsilon > 0$ defined in Assumption \ref{ass:local_control}, there exists a horizon $N^{\prime} \in \mathbb{N}$ such that for every initial state $x_{0} \in \mathbb{X}$, there is a time step $j \leq N^{\prime}$ and an admissible control sequence $u \in \mathbb{U}^{N^{\prime}}(x_{0})$ that satisfy $x_{u}(j, x_{0}) \in \mathcal{B}_{\varepsilon}(x^{e})$.\end{assumption}
\color{black}
\begin{theorem}[Adapted from \citet{GRUNE2014}]\label{theo:grune2014}
Consider the economic MPC problem without terminal conditions arising from the receding horizon solution to OCP \eqref{ocp_gen} satisfying Assumptions \ref{ass:dissi}-\ref{ass:finite_control}. Then, there exists a sufficiently large horizon $N_p \in \mathbb{N}$, such that the closed-loop system defined by \eqref{fg_ps_dis} is practically asymptotically stable w.r.t \textcolor{black}{$\epsilon$, where $\epsilon$ $\rightarrow 0$ as $N_p \rightarrow \infty$}.
\end{theorem}
\begin{proposition}[Adapted from \citet{Grune2017}]\label{prep:grune2017}
Let Assumptions \ref{ass:dissi}-\ref{ass:finite_control} hold. If, for the horizon $N_p \in \mathbb{N}$, the OCP \eqref{ocp_gen} is feasible for $j=0$ and $x_0\in \mathbb{X}$, then it is feasible for all $j \in \mathbb{N}$.
\end{proposition}
\color{black}
\begin{proposition}\label{prop_dotlambda}
Under Assumption \ref{ass:regularity}, for a sufficiently small sampling period $T_s > 0$, the continuous-time (limit-form) inequality
\small\begin{equation}
    \dot{\lambda}(x) \leq \ell(x,u) - \ell(x^e,u^e) - \alpha_\ell(\|x - x^e\|)\label{dotlambda}
\end{equation}
\normalsize
is sufficient to guarantee that the discrete-time strict dissipativity condition \eqref{dissipa:discrete} holds up to terms of order $\mathcal{O}(T_s)$.
\end{proposition}
\begin{proof}
By integrating the continuous-time strict dissipativity inequality \citep{willems1972, Angeli2012} over a single sampling period, and using the identity $\lambda(g_d(x,u)) - \lambda(x) = \lambda(\phi(t+T_s; t, x, u)) - \lambda(\phi(t; t, x, u))$, we have
\small$$\lambda(g_d(x,u)) - \lambda(x) \leq \int_t^{t+T_s} \left[ \ell_c(\phi(\tau),u) - \ell(x^e,u^e) - \alpha_\ell(\|\phi(\tau)-x^e\|) \right] d\tau.$$\normalsize
Evaluating this integral using the definition of the discrete-time cost $\ell(x,u)$ and the Lipschitz properties established in Assumption \ref{ass:regularity}, it follows that
\small\begin{equation}
\lambda(g_d(x,u)) - \lambda(x)
\leq T_s [\ell(x,u) - \ell(x^e,u^e) - \alpha_\ell(\|x - x^e\|)] + C_1 T_s^2,
\end{equation}
\normalsize
where $C_1$ is a constant capturing the discretization error. Dividing both sides by $T_s$ yields
\small\begin{equation}
    \frac{\lambda(g_d(x,u)) - \lambda(x)}{T_s}
    \leq \ell(x,u) - \ell(x^e,u^e) - \alpha_\ell(\|x - x^e\|) + \mathcal{O}(T_s).
\end{equation}
\normalsize
Finally, taking the limit as $T_s \to 0$, the left-hand side converges to the time derivative $\dot{\lambda}(x)$, thereby recovering the limit-form inequality \eqref{dotlambda}.
\end{proof}
\color{black}
\section{Nonlinear Model Predictive Control Schemes}\label{section:NMPC_form}
This section describes the proposed control strategies for \GR{motion control} a class of underactuated mechanical systems with second-order nonholonomic constraints modeled on $\mathsf{SE(3)}$. Both schemes regulate the vehicle to a desired equilibrium \GR{and} use an attitude error function defined on $\mathsf{SO(3)}$. The fundamental difference between these two NMPC schemes is that \GR{the second formulation incorporates mission time as an economic criterion, whereas the first one does not.}

\subsection{NMPC for motion control of a class of underactuated mechanical systems on $\mathsf{SE(3)}$}\label{subsection:scheme01}
Considering the discrete-time system $g_d$ given by \eqref{discrete01}-\eqref{eq:Rdyamics04}, the NMPC scheme repeatedly solves, in each sample time $k$, the following optimal control problem:
\small
\begin{gather}
\label{ocp_nmpc}
\begin{matrix}
	\displaystyle
	\min_{x,u}\sum_{j=0}^{N_p}\|\bar{x}(k+j)\|^{2}_{Q_{\bar{x}}}+ \|\bar{u}(k+j)\|^2_{Q_{\bar{u}}}\\  
	\begin{array}{rll}
	\textrm{s.t.} & x(j+1)=g_d(x(j),u(j)), & x_0=x(k),\\
	 &x \in \mathbb{X}, & \forall j\in \mathbb{N}_p,\\
	 & u \in\mathbb{U}, &  \forall j\in \mathbb{N}_p,
	\end{array} 
	\end{matrix}
\end{gather}
\normalsize
where $\bar{x}\coloneqq\matriz{(\xi^{\mathcal{I}})^\top & (v^{\mathcal{I}})^\top & \Psi(R^{\mathcal{I}}_{\mathcal{B}},R_d) & (\omega^{\mathcal{B}})^\top}^\top \in \mathbb{R}^{10}$ and $\bar{u}\coloneqq\matriz{\|f^{\mathcal{I}}\|& (\tau^{\mathcal{B}})^\top}^\top \in \mathbb{R}^4$, with $Q_{\bar{x}}>0$ being defined such that
$\|\bar{x}\|^{2}_{Q_{\bar{x}}}=k_p\|\xi^\mathcal{I}\|^{2}+k_v\|v^\mathcal{I}\|^{2}+k_\omega\|\omega^\mathcal{B}\|^{2}+k_R\Psi^2(R^{\mathcal{I}}_{\mathcal{B}})$, and $Q_{\bar{u}}>0$ being defined such that $\|\bar{u}\|^2_{Q_{\bar{u}}}=k_f\|f^{\mathcal{I}}\|^2+k_\tau \|\tau^\mathcal{B}\|^2$.
\begin{remark}
Without loss of generality, the position equilibrium is \GR{set} at the origin, $(\xi^{\mathcal{I}})^e=\matriz{0&0&0}^\top$, and \GR{\textcolor{black}{the equilibrium points for $R_d$, $v^{\mathcal{I}}_d$, and $\omega^{\mathcal{B}}_d$ are defined based on the system characteristics.} Therefore, at equilibrium, \textcolor{black}{$\ell(\bar{x}^e, \bar{u}^e) = 0$.}}

\end{remark}
\subsection{NMPC for fast motion control of a class of underactuated mechanical systems on $\mathsf{SE(3)}$}\label{subsection:scheme02}
Considering the discrete-time system, $g_d$, given by \eqref{discrete01}-\eqref{eq:Rdyamics04}, the economic NMPC scheme repeatedly solves, in each sample time $k$, the following optimal control problem:
\small
\begin{gather}
\label{ocp_tonmpc}
\begin{matrix}
	\displaystyle
	\min_{x,u}\sum_{j=0}^{N_p}\zeta^j\|\bar{x}(k+j)\|^{2}_{Q_{\bar{x}}}+ \|\bar{u}(k+j)\|^2_{Q_{\bar{u}}}\\  
	\begin{array}{rll}
	\textrm{s.t.} & x(j+1)=g_d(x(j),u(j)), & x_0=x(k),\\
	 &x \in \mathbb{X}, & \forall j\in \mathbb{N}_p,\\
	 & u \in\mathbb{U}, &  \forall j\in \mathbb{N}_p,
	\end{array} 
	\end{matrix}
\end{gather}
\normalsize
where $\zeta>1$ is an exponential weight in the stage cost \GR{to \VL{induce} 
rapid convergence to equilibrium} \citep{Verschueren2017}, while the other variables are defined as in the previous NMPC formulation. It is \GR{noteworthy} that the only difference between the OCPs \eqref{ocp_nmpc} and \eqref{ocp_tonmpc} is the exponential weight, $\zeta$.
\begin{remark}\label{remark:tonmpc}
\GR{According to Theorem 1 of \citet{Verschueren2017}, if $N_p$ is greater than or equal to the time-optimal solution, there exists a number $\zeta_1$ such that for all $\zeta > \zeta_1$, the solution of \eqref{ocp_tonmpc} becomes time-optimal. However, since the time-optimality in motion control problems mainly depends on the distance the vehicle must travel, which can be considerable, this work does not focus on the time-optimal solution. As a result, $\zeta$ is used as a tuning gain to promote faster motion.}
\end{remark}

\section{Feasibility and Stability Analysis}\label{section:stability}
This section provides theoretical results ensuring the proposed controllers' closed-loop stability and recursive feasibility.

\subsection{NMPC for motion control of a class of underactuated mechanical systems on $\mathsf{SE(3)}$}
Underactuated motion systems with second-order nonholonomic constraints represented by \eqref{ms1}-\eqref{ms4} are not linearly controllable by a continuous time-invariant feedback control law due to second-order nonholonomic constraints \citep{Brockett83}, making the conventional Kalman rank condition \citep{Kalman63} \GR{unsuitable for assessing controllability}. Instead, the $N$-controllability of the system is verified \citep{Ferramosca2014}.
\begin{definition}
A linearized discrete-time system defined by matrices $A$ and $B$ is said to be $N$-step controllable if the $N$-controllability matrix \GR{$C_{o_N} \coloneqq \matriz{A^{N-1}B & \cdots & AB & B}$} 
has rank greater \GR{than} or equal to the dimension of the system state vector $x\in\mathbb{R}^n$, i.e., $\textrm{rank}(C_{o_N})\geq n$. This means that there exists a sequence $\{u(0),u(1),\cdots,u(N-1)\}$ that can transfer the system from any $x_0$ to $x^e$ in N steps.
\end{definition}
\begin{lemma}\label{lemma:control}
\textcolor{black}{For a sufficiently small sampling period $T_s>0$,} the \textcolor{black}{discrete-time} system \textcolor{black}{resulting from} \eqref{ms1}-\eqref{ms4} is four-step controllable \textcolor{black}{at every $x \in \mathbb{X}$}.
\end{lemma}
\begin{proof}
Due to the state variable $R^{\mathcal{I}}_{\mathcal{B}} \in$ $\mathsf{SO(3)}$, the conventional Jacobian linearization method cannot be used. Thus, from the exponential map of $\mathsf{SO(3)}$ and the isomorphism between $\mathfrak{so}(3)$ and $\mathbb{R}^3$, the infinitesimal variation principle of the rotational motion is applied, satisfying \citep{Lee2018}
\small
\begin{align}
    \delta R^{\mathcal{I}}_{\mathcal{B}}&=R^{\mathcal{I}}_{\mathcal{B}}\mathcal{S}(\eta),\\
    \delta \omega^{\mathcal{B}} &= \mathcal{S}(\omega^{\mathcal{B}})\eta + \dot{\eta},
\end{align}
\normalsize
where a differentiable curve, $\eta \in \mathbb{R}^3$, is used to represent $\delta R^{\mathcal{I}}_{\mathcal{B}}$. Thus, 
the linearized equations of \eqref{ms1}-\eqref{ms4} in $\mathsf{TSE(3)}$ can be formulated as
\small
\begin{gather} \label{linerizationSE3}
\dot{\delta x}=A\delta x+B\delta u, 
\end{gather}
\normalsize
where $\delta x \coloneqq \matriz{(\delta \xi^{\mathcal{I}})^\top&(\delta v^{\mathcal{I}})^\top&\eta^\top &(\delta \omega^{\mathcal{B}})^\top}^\top \in \mathbb{R}^{12}$, and $\delta u\coloneqq\matriz{\delta T&(\delta \tau^{\mathcal{B}})^\top}^\top \in \mathbb{R}^{4}$. The matrices $A \in \mathbb{R}^{12\times 12}$ and $B \in \mathbb{R}^{12\times4}$ are computed as
\small
\[
A=\matriz{
{\textbf{0}_{3}}&{\textit{I}_3}&{\textbf{0}_{3}}&{\textbf{0}_{3}}\\
{\textbf{0}_{3}}&{\textbf{0}_{3}}&{-\frac{T}{m}R^{\mathcal{I}}_{\mathcal{B}}\mathcal{S}(e^{\mathcal{B}})}&{\textbf{0}_{3}}\\
{\textbf{0}_{3}}&{\textbf{0}_{3}}&{-\mathcal{S}(\omega^{\mathcal{B}})}&{\textit{I}_3}\\
{\textbf{0}_{3}}&{\textbf{0}_{3}}&{\textbf{0}_{3}}&{J^{-1}(\mathcal{S}(J\omega^{\mathcal{B}})-\mathcal{S}(\omega^{\mathcal{B}})J)}
},
\]
\[
B=\matriz{{\textbf{0}_{3\times1}}&{\textbf{0}_{3}}\\
{\frac{1}{m}R^{\mathcal{I}}_{\mathcal{B}}e^{\mathcal{B}}}&{\textbf{0}_{3}}\\
{\textbf{0}_{3\times1}}&{\textbf{0}_{3}}\\
{\textbf{0}_{3\times1}}&{J^{-1}}
}.
\]
\normalsize
\color{black}
The discrete-time model of \eqref{linerizationSE3} under zero-order hold is given by
\small
\begin{align}
\delta x(j+1) = A_d \delta x(j) + B_d \delta u(j),
\label{eq:l_model_disc}
\end{align}
\normalsize
assuming a sufficiently small sampling period $T_s > 0$. \color{black}

It can be verified that, considering $N=4$, \GR{$\textrm{rank}\left({\matriz{A_d^3B_d& A_d^2B_d& A_dB_d &B_d}}\right)=12$}, 
which is equal to the dimension of the state vector $\delta x$. Therefore, the system \eqref{ms1}-\eqref{ms4} is four step controllable \textcolor{black}{at every $x \in \mathbb{X}$.}
\end{proof}
\begin{remark} \textcolor{black}{According to \citet[Theorem 7]{Sontag1998}, the controllability of a linearized system implies Small-Time Local Controllability (STLC) for the corresponding nonlinear system. Consequently, Lemma \ref{lemma:control} establishes that the linearized controllability of our system is sufficient to satisfy the requirements of Assumption \ref{ass:local_control}.} \end{remark}
\color{black}
\begin{lemma}\label{lemma:finite_time}The system \eqref{fg_ps_dis} is finite-time controllable into $\mathcal{B}_\epsilon(x^e)$. Specifically, there exists a horizon $N' \in \mathbb{N}$ such that for every $x_0 \in \mathbb{X}$, there is a time step $j \leq N'$ and an admissible control sequence $u \in \mathbb{U}^{N'}(x_0)$ satisfying $x_u(j,x_0) \in \mathcal{B}_\epsilon(x^e)$.\end{lemma}

\begin{proof}Combining the result of Lemma \ref{lemma:control} with the compactness of $\mathbb{X}$ and the continuity of the linearization, the proof follows a three-step constructive argument: 1) Uniform Local Controllability - since the linearized system is $4$-step controllable at every point in $\mathbb{X}$ (Lemma \ref{lemma:control}) and the linearization matrices vary continuously over the compact set $\mathbb{X}$, there exists a uniform radius $\epsilon > 0$ such that the system satisfies the local controllability requirements of Assumption \ref{ass:local_control} within $\mathcal{B}_\epsilon(x)$ for every $x \in \mathbb{X}$; 2) Finite Covering - due to the compactness of $\mathbb{X}$, there exists a finit convering $\{\mathcal{B}_{\epsilon/2}(x_i)\}_{i=1}^M$; and 3) Finite-Step Reachability - for any initial state $x_0 \in \mathbb{X}$, a finite trajectory to $x^e$ can be constructed by chaining control sequences through these overlapping neighborhoods. Because each neighborhood is locally controllable in at most 4 steps, the total number of steps required to reach $\mathcal{B}_\epsilon(x^e)$ is finite and bounded by $N' = 4 \times (M + 1)$.
\end{proof}
\color{black}
\begin{lemma}\label{lemma:dissipa}
\textcolor{black}{For a sufficiently small sampling period $T_s > 0$,} if $k_vk_f\geq0.25$ and $k_\omega k_\tau \geq0.25$, then the OCP \eqref{ocp_nmpc} is strictly dissipative with respect to the equilibrium $\textcolor{black}{(\bar{x}^e,\bar{u}^e)} \in \mathbb{X} \times \mathbb{U}$.
\end{lemma}
\begin{proof}
Consider the continuous time storage function candidate $\lambda: \mathsf{TSE(3)}\rightarrow\mathbb{R}$, defined as
\small
\begin{gather}\label{lagrangian}
    \lambda(x)=\frac{1}{2}(\omega^{\mathcal{B}})^\top
    J\omega^{\mathcal{B}} +\frac{1}{2}\|v^{\mathcal{I}}\|^2-\mathcal{P}(R^{\mathcal{I}}_{\mathcal{B}},\xi^{\mathcal{I}}),
\end{gather}
\normalsize
which is the Lagrangian of the system on $\mathsf{TSE(3)}$, where $\mathcal{P}(R^{\mathcal{I}}_{\mathcal{B}},\xi^{\mathcal{I}})$ is the potential energy that depends on the configuration $(R^{\mathcal{I}}_{\mathcal{B}},\xi^{\mathcal{I}})\in $ $\mathsf{SE(3)}$. 
It can be \GR{shown that for a rigid body rotating and translating} in three dimensions under constant external forces and torques, the derivative of \eqref{lagrangian} \GR{represents} the mechanical power of the system, given by
\small
\begin{gather}\label{deriva:lambda}
    \frac{d}{dt}(\lambda(x))=(\omega^{\mathcal{B}})^\top\tau^{\mathcal{B}}+(v^{\mathcal{I}})^\top f^{\mathcal{I}}.
\end{gather}
\normalsize
\textcolor{black}{From Proposition \ref{prop_dotlambda}}, for $\ell(x,u)$ defined by OCP \eqref{ocp_nmpc}, with $\textcolor{black}{\ell(\bar{x}^e,\bar{u}^e)=0}$, and considering \eqref{deriva:lambda}, the inequality \eqref{dotlambda} becomes
\small
\begin{gather}\label{inequality_one}
     (\omega^{\mathcal{B}})^\top\tau^{\mathcal{B}}+(v^{\mathcal{I}})^\top f^{\mathcal{I}} \leq \|\bar{x}\|^{2}_{Q_{\bar{x}}}+ \|\bar{u}\|^2_{Q_{\bar{u}}} -\textcolor{black}{\alpha_\ell(\|\bar{x}-\bar{x}^e\|)}.
\end{gather}
\normalsize
The inequality \eqref{inequality_one} is equivalent to
\small
\begin{gather}\label{inequality_two}
    h_1(x)+h_2(x) \leq k_p\|\xi^\mathcal{I}\|^{2}+k_R(\Psi(R^{\mathcal{I}}_{\mathcal{B}}))^2 -\textcolor{black}{\alpha_\ell(\|\bar{x}-\bar{x}^e\|)},
\end{gather}
\normalsize
where \GR{$h_1(x)=(v^{\mathcal{I}})^\top f^{\mathcal{I}} -k_v\|v^\mathcal{I}\|^{2}-k_f\|f^{\mathcal{I}}\|^2$ and $h_2(x) = (\omega^{\mathcal{B}})^\top\tau^{\mathcal{B}}-k_\omega\|\omega^\mathcal{B}\|^{2}-k_\tau \|\tau^\mathcal{B}\|^2$.} 

For convenience, \eqref{inequality_two} will be analyzed by parts. First, by using the basic concept of quadratic difference between two real numbers, 
\GR{yields $\left({\sqrt{k_v}\|v^\mathcal{I}\|-\sqrt{k_f}\|f^{\mathcal{I}}\|}\right)^2\geq0,$}
which can be rewritten as \GR{$2\sqrt{k_vk_f}\|v^\mathcal{I}\|\|f^{\mathcal{I}}\|-k_v\|v^\mathcal{I}\|^{2}-k_f\|f^{\mathcal{I}}\|^2\leq0.$} 
Additionally, the Cauchy-Schwarz inequality  \GR{give us $(v^\mathcal{I})^\top f^{\mathcal{I}}\leq \|v^\mathcal{I}\|\|f^{\mathcal{I}}\|$.} 
Thus, if $k_vk_f\geq0.25$, then
\small
\begin{multline*}
\GR{    (v^\mathcal{I})^\top f^{\mathcal{I}}\leq k_v\|v^\mathcal{I}\|^{2}+k_f\|f^{\mathcal{I}}\|^2\numberthis\label{ine:h1} \implies h_1(x)\leq 0~\forall~ x \in \mathsf{TSE(3)}.}
\end{multline*}
\normalsize
Second, by following a similar approach, it can be verified that, for $k_\omega k_\tau \geq0.25$, $h_2(x)\leq 0~\forall~ x \in \textrm{TSE(3)}$. Thus, \GR{$h_1(x)+h_2(x) \leq 0 ~\forall~ x \in \mathsf{TSE(3)}$.}
Finally, if $k_vk_f\geq0.25$, $k_\omega k_\tau \geq0.25$, and \textcolor{black}{since there exists} a function $\alpha_\ell \in \mathcal{K}_\infty$ such that \GR{$k_p\|\xi^\mathcal{I}\|^{2}+k_R(\Psi(R^{\mathcal{I}}_{\mathcal{B}}))^2 \geq\textcolor{black}{\alpha_\ell(\|\bar{x}-\bar{x}^e\|)}$,} 
then, the inequality \eqref{inequality_two} holds, and, therefore, the OCP \eqref{ocp_nmpc} is strictly dissipative with respect to the equilibrium $(x^e,u^e)$.
\end{proof}

The following theorem, based on the main dissipativity-based stability result for MPC schemes without terminal conditions by \citet{GRUNE2014}, establishes a minimum stabilizing horizon for the OCP \eqref{ocp_nmpc}.
\begin{theorem}
The NMPC scheme defined by the OCP \eqref{ocp_nmpc} is practically asymptotically stable for a sufficiently large horizon $N_p \in \mathcal{N}\coloneqq\{n_p \in \mathbb{N}:n_p\geq4\}$.
\end{theorem}
\color{black}
\begin{proof}
Lemmas \ref{lemma:control}, \ref{lemma:finite_time}, and \ref{lemma:dissipa}, together with Assumption \ref{ass:regularity}, ensure that all conditions of Theorem \ref{theo:grune2014} are satisfied. Therefore, the NMPC scheme is practically asymptotically stable for $N_p \geq 4$.
\end{proof}
\color{black}
\begin{proposition}
Assuming that, for horizon $N_p\in\mathcal{N}$, the OCP \eqref{ocp_nmpc} is feasible for $k=0$, then it is recursively feasible, i.e., it is feasible for all $k \in \mathbb{N}$.
\end{proposition}
\begin{proof}
The recursive feasibility follows from Proposition \ref{prep:grune2017}.
\end{proof}

\subsection{NMPC for fast motion control of a class of underactuated mechanical systems on $\mathsf{SE(3)}$}
\GR{This subsection evaluates} the closed-loop stability and recursive feasibility of the NMPC scheme described in Section \ref{subsection:scheme02}. First, the following lemma specifies the conditions under which the OCP \eqref{ocp_tonmpc} is strictly dissipative.
\begin{lemma}\label{lemma:tonmpc}
\textcolor{black}{For a sufficiently small sampling period $T_s > 0$,} if $k_vk_f\geq0.25$ and $k_\omega k_\tau \geq0.25$, then the OCP \eqref{ocp_tonmpc} is strictly dissipative with respect to the equilibrium $\textcolor{black}{(\bar{x}^e,\bar{u}^e)} \in \mathbb{X} \times \mathbb{U}$.
\end{lemma}
\begin{proof}
Consider \eqref{lagrangian} as a continuous time storage function candidate. The dissipation inequality \eqref{dotlambda} considering OCP \eqref{ocp_tonmpc} yields
\small
\begin{gather}\label{inequality_one_TONMPC}
     (\omega^{\mathcal{B}})^\top\tau^{\mathcal{B}}+(v^{\mathcal{I}})^\top f^{\mathcal{I}} \leq \zeta^t\|\bar{x}\|^{2}_{Q_{\bar{x}}}+ \|\bar{u}\|^2_{Q_{\bar{u}}} -\textcolor{black}{\alpha_\ell(\|\bar{x}-\bar{x}^e\|)}.
\end{gather}
\normalsize
The inequality \eqref{inequality_one_TONMPC} can be rewritten as
\small
\begin{gather}\label{inequality_two_TONMPC}
    h_3(x)+h_4(x) \leq \zeta^t(k_p\|\xi^\mathcal{I}\|^{2}+k_R(\Psi(R^{\mathcal{I}}_{\mathcal{B}}))^2) -\textcolor{black}{\alpha_\ell(\|\bar{x}-\bar{x}^e\|)},
\end{gather}
\normalsize
where \GR{$h_3(x)=(v^{\mathcal{I}})^\top f^{\mathcal{I}} -\zeta^tk_v\|v^\mathcal{I}\|^{2}-k_f\|f^{\mathcal{I}}\|^2$ and $h_4(x) = (\omega^{\mathcal{B}})^\top\tau^{\mathcal{B}}-\zeta^tk_\omega\|\omega^\mathcal{B}\|^{2}-k_\tau \|\tau^\mathcal{B}\|^2$.} 

From \eqref{ine:h1}, if $k_vk_f\geq0.25$, then \GR{$(v^\mathcal{I})^\top f^{\mathcal{I}}\leq k_v\|v^\mathcal{I}\|^{2}+k_f\|f^{\mathcal{I}}\|^2$.} 
Since, for $\zeta>1$, $\zeta^t \in \mathcal{K}_\infty$, then \GR{$(v^\mathcal{I})^\top f^{\mathcal{I}}\leq \zeta^tk_v\|v^\mathcal{I}\|^{2}+k_f\|f^{\mathcal{I}}\|^2 \implies h_3(x)\leq 0~\forall~ x \in \mathsf{TSE(3)}$.} 
%
Following the same steps, it can be verified that, for $k_\omega k_\tau \geq0.25$, $h_4(x)\leq 0~\forall~ x \in \textrm{TSE(3)}$. Thus, \GR{$h_3(x)+h_4(x) \leq 0 ~\forall~ x \in \mathsf{TSE(3)}$.} 
Finally, if $k_vk_f\geq0.25$, $k_\omega k_\tau \geq0.25$, and \textcolor{black}{since there exists} a function $\alpha_\ell \in \mathcal{K}_\infty$ such that \GR{$\zeta^t(k_p\|\xi^\mathcal{I}\|^{2}+k_R(\Psi(R^{\mathcal{I}}_{\mathcal{B}}))^2) \geq\textcolor{black}{\alpha_\ell(\|\bar{x}-\bar{x}^e\|)}$,} 
\normalsize
then, the inequality \eqref{inequality_two_TONMPC} holds, and, therefore, the OCP \eqref{ocp_tonmpc} is strictly dissipative with respect to the equilibrium $\textcolor{black}{(\bar{x}^e,\bar{u}^e)}$.
\end{proof}
The following theorem specifies a minimum stabilizing horizon for OCP \eqref{ocp_tonmpc}.
\begin{theorem}
The NMPC scheme defined by the OCP \eqref{ocp_tonmpc} is practically asymptotically stable for a sufficiently large horizon $N_p \in \mathcal{N}$.
\end{theorem}
\color{black}
\begin{proof}
Lemmas \ref{lemma:control}, \ref{lemma:finite_time}, and \ref{lemma:tonmpc}, together with Assumption \ref{ass:regularity}, ensure that all conditions of Theorem \ref{theo:grune2014} are satisfied. Therefore, the NMPC scheme is practically asymptotically stable for $N_p \geq 4$.
\end{proof}
\color{black}
\begin{proposition}
\GR{Assuming the OCP \eqref{ocp_tonmpc} is feasible for $k=0$ with a horizon $N_p\in\mathcal{N}$}, then it is recursively feasible, i.e., it is feasible for all $k \in \mathbb{N}$.
\end{proposition}
\begin{proof}
The recursive feasibility follows from Proposition \ref{prep:grune2017}.
\end{proof}

\section{Numerical Example: Quadrotor UAV}\label{section:experiment}
In this section, the proposed NMPC schemes are corroborated using a quadrotor UAV in numerical experiments conducted within the ProVANT simulator\footnote{\url{https://github.com/Guiraffo/ProVANT-Simulator}}, a virtual environment based on Gazebo \citep{Koenig2004} and Robot Operating System (ROS) \citep{Quigley2009}. The experiments aim to achieve point-to-point motion, guiding the vehicle from an initial configuration $x_0$ to a desired set-point $x^e$. First, the NMPC scheme \eqref{ocp_nmpc} is tested with four different initial configurations. In the second part, the NMPC scheme \eqref{ocp_tonmpc} is evaluated using these configurations to assess not only controller efficiency but also the reduction in mission time.

\subsection{Experiment setting}
\GR{All experiments consider the quadrotor UAV model defined by \eqref{ms1}-\eqref{ms4}, with $e^{\mathcal{B}}=e_3$ and $f_e^{\mathcal{I}}=-a_ge_3$, where $e_3=\matriz{0&0&1}^\top$ and $a_g$ represents gravitational acceleration. To account for the limited thrust of each propeller, the following expression maps each thrust value to the input vector:}
\small
\begin{gather}
u=\matriz{ 
{T}\\ 
{\tau_{\phi}}\\ 
{\tau_{\theta}}\\
{\tau_{\psi}}
} =\matriz{ 
{1} & {1} & {1} & {1} \\ 
{0} & {l} & {0} & {-l} \\ 
{-l} & {0} & {l} & {0}\\ 
{c_{\tau f}} & {-c_{\tau f}} & {c_{\tau f}} & {-c_{\tau f}} 
}
\matriz{ 
{f_1}\\ 
{f_2} \\ 
{f_3}\\
{f_4}
},\label{input_chap4}
\end{gather}
\normalsize
where $l$ is the distance from the quadrotor's center of mass to \GR{each rotor's center}, and $c_{\tau f}$ is the constant \GR{determining the torque produced by} each propeller. The quadrotor UAV parameters were obtained from \citet{raffotese} \textcolor{black}{ (see Table 2.1 on page 45 for details).} 
The controller runs at 100 Hz, \GR{with a prediction horizon of $N_p=40$ for all results presented in the following subsection. The equilibrium set-point is defined as $x^e=(\textit{I}_3,\matriz{0&0&4}^\top, 0_{3\times1}, 0_{3\times1})$. Based on} Lemmas \ref{lemma:dissipa} and \ref{lemma:tonmpc}, the gains for both schemes are tuned as $k_p=150$, $k_v=30$, $k_R=10$, $k_1=10$, $k_3=1$, $k_\omega=0.85$, $k_f=5e^{-2}$,  $k_\tau=0.3$, and $\zeta=1.2$. The initial configurations are given by $x_{01}=(R_{01},\xi_{01}, 0_{3\times1}, 0_{3\times1})$, $x_{02}=(R_{02},\xi_{02}, 0_{3\times1}, 0_{3\times1})$, $x_{03}=(R_{03},\xi_{03}, 0_{3\times1}, 0_{3\times1})$, and $x_{04}=(R_{04},\xi_{04}, 0_{3\times1}, 0_{3\times1})$, where
\small
\[
R_{01}=\matriz{
    0.1543&0.9880&0 \\
    0.1954&-0.0305&-0.9802 \\
   -0.9685&0.1512&-0.1978

}, ~~~ \xi_{01}=\matriz{
  4\\
     5\\
     7
},
\]
\[
R_{02}=\matriz{
  -0.3045 &   0.2837  &  0.9093 \\
   -0.4447  &  0.8019 &  -0.3990\\
   -0.8424  & -0.5258  & -0.1180

}, ~~~ \xi_{02}=\matriz{
  -4\\
     -5\\
     2
},
\]
\[
R_{03}=\matriz{
    0.5175  & -0.3541  & -0.7790 \\
    0.8168  &  0.4755  &  0.3266 \\
    0.2548  & -0.8053 &   0.5353

}, ~~~ \xi_{03}=\matriz{
  -4\\
     5\\
     7
},
\]
\[
R_{04}=\matriz{
    0.8660 &  -0.5000   & 0 \\
   -0.5000 &  -0.8660  &  0\\
    0 &  0  & -1

}, ~~~ \xi_{04}=\matriz{
  3\\
     -4\\
     9
}.
\]
\normalsize
\begin{remark}
The prediction horizon $N_p$ is \GR{chosen based on} a trade-off between computational cost and control performance. \GR{For instance, controllers can be tuned with $N_p=4$; however, a short stabilizing prediction horizon requires more} conservative vehicle behavior, resulting in slower and smoother rotational and translational movements.
\end{remark}

\GR{The OCP solution employs the direct method introduced in \citet{bock84} to evaluate solutions for \eqref{ocp_nmpc} and \eqref{ocp_tonmpc} at each sample time. This method parameterizes infinite-dimensional decision variables to approximate the original OCP as a finite-dimensional nonlinear program (NLP) \citep{RawlingsMPC}. The direct multiple-shooting technique is used, where both piecewise control discretization and states at interval boundaries are decision variables in a finite-dimensional optimization problem.} The CasADI toolbox \citep{Andersson2018} solves this NLP, interfacing with the IPOPT solver \citep{Andreas2002}. The experiments are conducted on a general-purpose notebook with an Intel Core i7 4500U processor (2 physical and 4 logical cores at 1.8 GHz), 8 GB of RAM, and an NVidia GT 750M GPU, running Ubuntu Linux version 18.04.

\subsection{Simulation results}
 
\begin{figure}[htb!]
	\centering
\def\svgwidth{1\columnwidth}
		\scriptsize{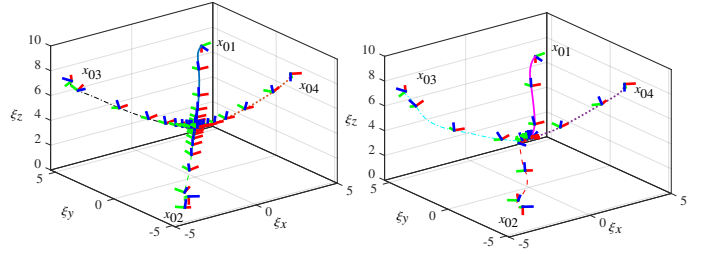}
		\caption{The UAV trajectories obtained from the experiments: the NMPC scheme $\eqref{ocp_nmpc}$ on the left and the NMPC scheme $\eqref{ocp_tonmpc}$ on the right.}\label{figura7}
\end{figure}

\begin{figure}[htb!]
	\centering
\def\svgwidth{1\columnwidth}
		\scriptsize{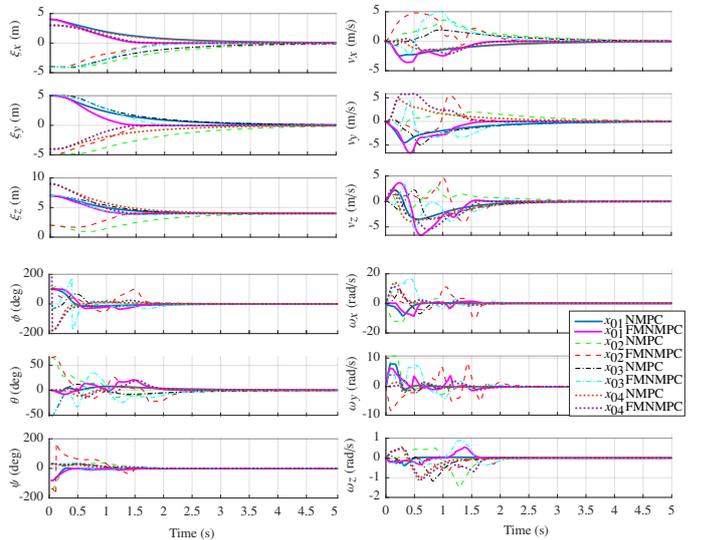}
		\caption{\textcolor{black}{States trajectories of the UAV during numerical experiments.}}\label{figura1}
\end{figure}

\GR{Figure \ref{figura7} shows the UAV trajectories during \textcolor{black}{numerical} experiments\footnote{A video of the results is available at \href{https://youtu.be/mjKZvK4KNcY}{{https://youtu.be/mjKZvK4KNcY}}.}. Both sets of experiments use the same parameter values, including gains, prediction horizon, initial conditions, and equilibrium set-point. Four distinct initial conditions are considered, including a non-equilibrium configuration, for instance, where the UAV starts upside-down initial condition $x_{04}$. Controllers utilizing non-singularity-free parameterization struggle with such maneuvers. Additionally, stabilizing the UAV is particularly challenging as only positive thrust values can be applied.}
\GR{Figure \ref{figura1} demonstrates that both NMPC schemes successfully reach the equilibrium point $x^e$ for all initial conditions. The UAV attitude is represented using Euler angles in degrees for clarity. In all cases, the UAV attitude stabilizes at $I_3 \in$ $\mathsf{SO(3)}$, corresponding to zero degrees for roll, pitch, and yaw angles. Notably, the UAV reaches the origin more quickly with the fast motion NMPC scheme \eqref{ocp_tonmpc}, referred to as FMNMPC. Although the exponential weight in the stage cost encourages rapid stabilization of all states, the system's underactuation requires prioritizing certain states. This prioritization is achieved through the definition of controller gains. In these experiments, the faster behavior in the translational movement is evident, as the position gain, $k_p$, is significantly larger than the other gains. Consequently, using FMNMPC results in a considerably reduced settling time for the position vector $\xi$. Additionally, the vehicle exhibits more aggressive motion under the FMNMPC scheme, with larger angular excursions and higher maximum velocities.}

\begin{figure}[htb!]
	\centering
\def\svgwidth{0.8\columnwidth}
		\scriptsize{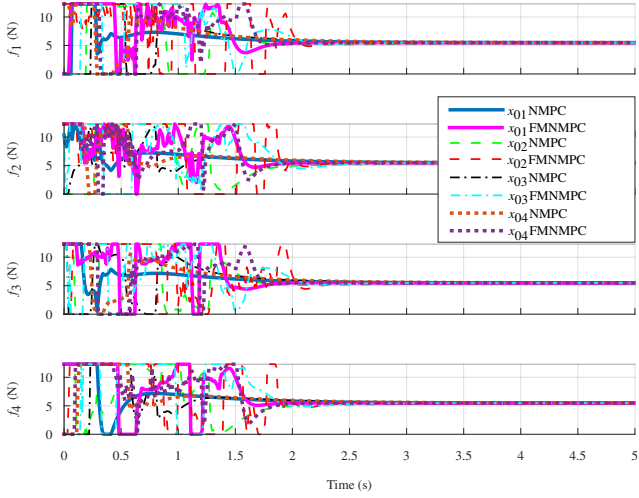}
		\caption{UAV propeller forces. The controller correctly respects the input bounds.}\label{figura5}
\end{figure}

The proposed NMPC schemes complied with all system constraints considered in the control design. Figure \ref{figura5} shows the propeller forces which are mapped to the input signals\GR{, showing that the input limits, \textcolor{black}{bounded between 0 and 12.3 N}, are satisfied by the controllers. This demonstrates the controllers' capability to handle constrained control problems.} Furthermore, the FMNMPC reaches the input limits more frequently than the NMPC scheme \eqref{ocp_nmpc}, indicating its aggressiveness.

\begin{figure}[htb!]
	\centering
\def\svgwidth{1\columnwidth}
		\scriptsize{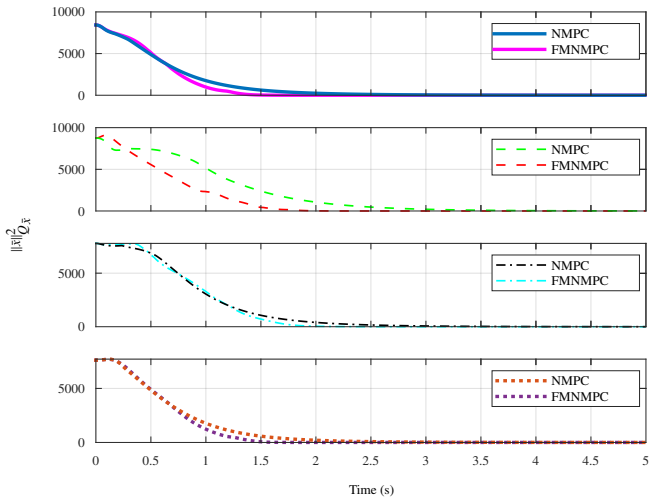}
		\caption{Variation of $\|\bar{x}\|^{2}_{Q_{\bar{x}}}$ and $\|\xi^{\mathcal{I}}\|$ for each initial condition and each proposed NMPC scheme.}\label{figura9}
\end{figure}

\GR{Lastly, Figure \ref{figura9} evaluate the influence of the exponential weight on the stage cost. Figure \ref{figura9} shows that the index $\|\bar{x}\|^{2}_{Q_{\bar{x}}}$ for FMNMPC exhibits a quicker decay in all scenarios, as expected, due to the exponential weight encouraging $\bar{x}$ to approach the equilibrium point rapidly at each prediction horizon. Additionally, Figure \ref{figura9} emphasizes the faster behavior in the translational movement, with an average settling time for FMNMPC being $48\%$ shorter, indicating a reduction in mission time. As noted in Remark \ref{remark:tonmpc}, the parameter $\zeta$ serves as a tuning gain, allowing for adjustments in the OCP \eqref{ocp_tonmpc} to further reduce settling time with a larger $\zeta$.}

\section{Conclusions}\label{section:conclusions}
\GR{This paper presented two practically asymptotically stable NMPC schemes without terminal conditions for controlling a class of underactuated mechanical systems with second-order nonholonomic constraints. The controllers were formulated on the $\mathsf{SE(3)}$ manifold, enabling a global and unique representation of vehicle motion. This singularity-free feature was evidenced in experiments where the quadrotor UAV successfully recovered from an upside-down position without additional controllers or optimization layers. The closed-loop stability and recursive feasibility were thoroughly assessed using recent theories on strict dissipativity and MPC schemes without terminal conditions. Furthermore, the proposed fast-motion NMPC scheme guided the vehicle to a desired equilibrium point while minimizing mission time as an economic criterion. Numerical experiments validated the theoretical results, showing that both controllers effectively reached the equilibrium point from various initial conditions. The analysis also explored the impact of the exponential weight on closed-loop performance, highlighting the ability to achieve a fast and stable system response with simple tuning parameters.} 

\GR{Future work will focus on extending the proposed NMPC schemes to address autonomous navigation in obstructed environments, integrating collision avoidance strategies, and analyzing the impact of disturbances on stability and economic criteria. Additionally, we aim to obtain experimental results by applying the proposed NMPC schemes to real-world underactuated mechanical systems.}



\bibliography{References}

\end{document}

%% file: Figuras/figure_trajectories.pdf_tex
\tiny
\begingroup%
  \makeatletter%
  \providecommand\color[2][]{%
    \errmessage{(Inkscape) Color is used for the text in Inkscape, but the package 'color.sty' is not loaded}%
    \renewcommand\color[2][]{}%
  }%
  \providecommand\transparent[1]{%
    \errmessage{(Inkscape) Transparency is used (non-zero) for the text in Inkscape, but the package 'transparent.sty' is not loaded}%
    \renewcommand\transparent[1]{}%
  }%
  \providecommand\rotatebox[2]{#2}%
  \newcommand*\fsize{\dimexpr\f@size pt\relax}%
  \newcommand*\lineheight[1]{\fontsize{\fsize}{#1\fsize}\selectfont}%
  \ifx\svgwidth\undefined%
    \setlength{\unitlength}{764.25224304bp}%
    \ifx\svgscale\undefined%
      \relax%
    \else%
      \setlength{\unitlength}{\unitlength * \real{\svgscale}}%
    \fi%
  \else%
    \setlength{\unitlength}{\svgwidth}%
  \fi%
  \global\let\svgwidth\undefined%
  \global\let\svgscale\undefined%
  \makeatother%
  \begin{picture}(1,0.3690721)%
    \lineheight{1}%
    \setlength\tabcolsep{0pt}%
    \put(0,0){\includegraphics[width=\unitlength,page=1]{figure_trajectories.pdf}}%
    \put(0.25612115,0.01574625){\color[rgb]{0.14901961,0.14901961,0.14901961}\makebox(0,0)[lt]{\lineheight{1.25}\smash{\begin{tabular}[t]{l}-5\end{tabular}}}}%
    \put(0.3766154,0.04836929){\color[rgb]{0.14901961,0.14901961,0.14901961}\makebox(0,0)[lt]{\lineheight{1.25}\smash{\begin{tabular}[t]{l}0\end{tabular}}}}%
    \put(0.49710964,0.08099237){\color[rgb]{0.14901961,0.14901961,0.14901961}\makebox(0,0)[lt]{\lineheight{1.25}\smash{\begin{tabular}[t]{l}5\end{tabular}}}}%
    \put(0,0){\includegraphics[width=\unitlength,page=2]{figure_trajectories.pdf}}%
    \put(0.22031063,0.01900826){\color[rgb]{0.14901961,0.14901961,0.14901961}\makebox(0,0)[rt]{\lineheight{1.25}\smash{\begin{tabular}[t]{r}-5\end{tabular}}}}%
    \put(0.13333738,0.05900028){\color[rgb]{0.14901961,0.14901961,0.14901961}\makebox(0,0)[rt]{\lineheight{1.25}\smash{\begin{tabular}[t]{r}0\end{tabular}}}}%
    \put(0.04636512,0.09899328){\color[rgb]{0.14901961,0.14901961,0.14901961}\makebox(0,0)[rt]{\lineheight{1.25}\smash{\begin{tabular}[t]{r}5\end{tabular}}}}%
    \put(0,0){\includegraphics[width=\unitlength,page=3]{figure_trajectories.pdf}}%
    \put(0.0434662,0.12059676){\color[rgb]{0.14901961,0.14901961,0.14901961}\makebox(0,0)[rt]{\lineheight{1.25}\smash{\begin{tabular}[t]{r}0\end{tabular}}}}%
    \put(0.0434662,0.15772521){\color[rgb]{0.14901961,0.14901961,0.14901961}\makebox(0,0)[rt]{\lineheight{1.25}\smash{\begin{tabular}[t]{r}2\end{tabular}}}}%
    \put(0.0434662,0.19485268){\color[rgb]{0.14901961,0.14901961,0.14901961}\makebox(0,0)[rt]{\lineheight{1.25}\smash{\begin{tabular}[t]{r}4\end{tabular}}}}%
    \put(0.0434662,0.23198014){\color[rgb]{0.14901961,0.14901961,0.14901961}\makebox(0,0)[rt]{\lineheight{1.25}\smash{\begin{tabular}[t]{r}6\end{tabular}}}}%
    \put(0.0434662,0.26910859){\color[rgb]{0.14901961,0.14901961,0.14901961}\makebox(0,0)[rt]{\lineheight{1.25}\smash{\begin{tabular}[t]{r}8\end{tabular}}}}%
    \put(0.0434662,0.30623606){\color[rgb]{0.14901961,0.14901961,0.14901961}\makebox(0,0)[rt]{\lineheight{1.25}\smash{\begin{tabular}[t]{r}10\end{tabular}}}}%
    \put(0,0){\includegraphics[width=\unitlength,page=4]{figure_trajectories.pdf}}%
    \put(-0.00130336,0.19556367){\makebox(0,0)[lt]{\lineheight{1.25}\smash{\begin{tabular}[t]{l}$\xi_z$\end{tabular}}}}%
    \put(0.07400083,0.04971709){\makebox(0,0)[lt]{\lineheight{1.25}\smash{\begin{tabular}[t]{l}$\xi_y$\end{tabular}}}}%
    \put(0.3897926,0.03396189){\makebox(0,0)[lt]{\lineheight{1.25}\smash{\begin{tabular}[t]{l}$\xi_x$\end{tabular}}}}%
    \put(0.30390983,0.30275228){\makebox(0,0)[lt]{\lineheight{1.25}\smash{\begin{tabular}[t]{l}$x_{01}$\end{tabular}}}}%
    \put(0.22506913,0.04597865){\makebox(0,0)[lt]{\lineheight{1.25}\smash{\begin{tabular}[t]{l}$x_{02}$\end{tabular}}}}%
    \put(0.10383097,0.26073349){\makebox(0,0)[lt]{\lineheight{1.25}\smash{\begin{tabular}[t]{l}$x_{03}$\end{tabular}}}}%
    \put(0.42666282,0.23973787){\makebox(0,0)[lt]{\lineheight{1.25}\smash{\begin{tabular}[t]{l}$x_{04}$\end{tabular}}}}%
    \put(0,0){\includegraphics[width=\unitlength,page=5]{figure_trajectories.pdf}}%
    \put(0.75229916,0.00030095){\color[rgb]{0.14901961,0.14901961,0.14901961}\makebox(0,0)[lt]{\lineheight{1.25}\smash{\begin{tabular}[t]{l}-5\end{tabular}}}}%
    \put(0.8727934,0.032924){\color[rgb]{0.14901961,0.14901961,0.14901961}\makebox(0,0)[lt]{\lineheight{1.25}\smash{\begin{tabular}[t]{l}0\end{tabular}}}}%
    \put(0.99328759,0.06554707){\color[rgb]{0.14901961,0.14901961,0.14901961}\makebox(0,0)[lt]{\lineheight{1.25}\smash{\begin{tabular}[t]{l}5\end{tabular}}}}%
    \put(0,0){\includegraphics[width=\unitlength,page=6]{figure_trajectories.pdf}}%
    \put(0.71648864,0.00356297){\color[rgb]{0.14901961,0.14901961,0.14901961}\makebox(0,0)[rt]{\lineheight{1.25}\smash{\begin{tabular}[t]{r}-5\end{tabular}}}}%
    \put(0.62951541,0.04355498){\color[rgb]{0.14901961,0.14901961,0.14901961}\makebox(0,0)[rt]{\lineheight{1.25}\smash{\begin{tabular}[t]{r}0\end{tabular}}}}%
    \put(0.54254315,0.08354799){\color[rgb]{0.14901961,0.14901961,0.14901961}\makebox(0,0)[rt]{\lineheight{1.25}\smash{\begin{tabular}[t]{r}5\end{tabular}}}}%
    \put(0,0){\includegraphics[width=\unitlength,page=7]{figure_trajectories.pdf}}%
    \put(0.53964419,0.10515147){\color[rgb]{0.14901961,0.14901961,0.14901961}\makebox(0,0)[rt]{\lineheight{1.25}\smash{\begin{tabular}[t]{r}0\end{tabular}}}}%
    \put(0.53964419,0.14227993){\color[rgb]{0.14901961,0.14901961,0.14901961}\makebox(0,0)[rt]{\lineheight{1.25}\smash{\begin{tabular}[t]{r}2\end{tabular}}}}%
    \put(0.53964419,0.1794074){\color[rgb]{0.14901961,0.14901961,0.14901961}\makebox(0,0)[rt]{\lineheight{1.25}\smash{\begin{tabular}[t]{r}4\end{tabular}}}}%
    \put(0.53964419,0.21653485){\color[rgb]{0.14901961,0.14901961,0.14901961}\makebox(0,0)[rt]{\lineheight{1.25}\smash{\begin{tabular}[t]{r}6\end{tabular}}}}%
    \put(0.53964419,0.25366332){\color[rgb]{0.14901961,0.14901961,0.14901961}\makebox(0,0)[rt]{\lineheight{1.25}\smash{\begin{tabular}[t]{r}8\end{tabular}}}}%
    \put(0.53964419,0.29079079){\color[rgb]{0.14901961,0.14901961,0.14901961}\makebox(0,0)[rt]{\lineheight{1.25}\smash{\begin{tabular}[t]{r}10\end{tabular}}}}%
    \put(0,0){\includegraphics[width=\unitlength,page=8]{figure_trajectories.pdf}}%
    \put(0.49611482,0.18809426){\makebox(0,0)[lt]{\lineheight{1.25}\smash{\begin{tabular}[t]{l}$\xi_z$\end{tabular}}}}%
    \put(0.57141904,0.04224767){\makebox(0,0)[lt]{\lineheight{1.25}\smash{\begin{tabular}[t]{l}$\xi_y$\end{tabular}}}}%
    \put(0.88721078,0.02649247){\makebox(0,0)[lt]{\lineheight{1.25}\smash{\begin{tabular}[t]{l}$\xi_x$\end{tabular}}}}%
    \put(0.80132803,0.29528287){\makebox(0,0)[lt]{\lineheight{1.25}\smash{\begin{tabular}[t]{l}$x_{01}$\end{tabular}}}}%
    \put(0.72248732,0.03850924){\makebox(0,0)[lt]{\lineheight{1.25}\smash{\begin{tabular}[t]{l}$x_{02}$\end{tabular}}}}%
    \put(0.60124915,0.25326408){\makebox(0,0)[lt]{\lineheight{1.25}\smash{\begin{tabular}[t]{l}$x_{03}$\end{tabular}}}}%
    \put(0.92408099,0.23226846){\makebox(0,0)[lt]{\lineheight{1.25}\smash{\begin{tabular}[t]{l}$x_{04}$\end{tabular}}}}%
  \end{picture}%
\endgroup%

%% file: Figuras/figure_states.pdf_tex
\tiny
\begingroup%
  \makeatletter%
  \providecommand\color[2][]{%
    \errmessage{(Inkscape) Color is used for the text in Inkscape, but the package 'color.sty' is not loaded}%
    \renewcommand\color[2][]{}%
  }%
  \providecommand\transparent[1]{%
    \errmessage{(Inkscape) Transparency is used (non-zero) for the text in Inkscape, but the package 'transparent.sty' is not loaded}%
    \renewcommand\transparent[1]{}%
  }%
  \providecommand\rotatebox[2]{#2}%
  \newcommand*\fsize{\dimexpr\f@size pt\relax}%
  \newcommand*\lineheight[1]{\fontsize{\fsize}{#1\fsize}\selectfont}%
  \ifx\svgwidth\undefined%
    \setlength{\unitlength}{766.64300537bp}%
    \ifx\svgscale\undefined%
      \relax%
    \else%
      \setlength{\unitlength}{\unitlength * \real{\svgscale}}%
    \fi%
  \else%
    \setlength{\unitlength}{\svgwidth}%
  \fi%
  \global\let\svgwidth\undefined%
  \global\let\svgscale\undefined%
  \makeatother%
  \begin{picture}(1,0.79356022)%
    \lineheight{1}%
    \setlength\tabcolsep{0pt}%
    \put(0,0){\includegraphics[width=\unitlength,page=1]{figure_states.pdf}}%
    \put(0.54698574,0.18175309){\color[rgb]{0.14901961,0.14901961,0.14901961}\makebox(0,0)[rt]{\lineheight{1.25}\smash{\begin{tabular}[t]{r}-10\end{tabular}}}}%
    \put(0.54698574,0.22455237){\color[rgb]{0.14901961,0.14901961,0.14901961}\makebox(0,0)[rt]{\lineheight{1.25}\smash{\begin{tabular}[t]{r}0\end{tabular}}}}%
    \put(0.54698574,0.2673526){\color[rgb]{0.14901961,0.14901961,0.14901961}\makebox(0,0)[rt]{\lineheight{1.25}\smash{\begin{tabular}[t]{r}10\end{tabular}}}}%
    \put(0,0){\includegraphics[width=\unitlength,page=2]{figure_states.pdf}}%
    \put(0.54698574,0.30486613){\color[rgb]{0.14901961,0.14901961,0.14901961}\makebox(0,0)[rt]{\lineheight{1.25}\smash{\begin{tabular}[t]{r}-20\end{tabular}}}}%
    \put(0.54698574,0.3491876){\color[rgb]{0.14901961,0.14901961,0.14901961}\makebox(0,0)[rt]{\lineheight{1.25}\smash{\begin{tabular}[t]{r}0\end{tabular}}}}%
    \put(0.54698574,0.39350809){\color[rgb]{0.14901961,0.14901961,0.14901961}\makebox(0,0)[rt]{\lineheight{1.25}\smash{\begin{tabular}[t]{r}20\end{tabular}}}}%
    \put(0,0){\includegraphics[width=\unitlength,page=3]{figure_states.pdf}}%
    \put(0.76385329,0.00738441){\color[rgb]{0.14901961,0.14901961,0.14901961}\makebox(0,0)[t]{\lineheight{1.25}\smash{\begin{tabular}[t]{c}Time (s)\end{tabular}}}}%
    \put(0,0){\includegraphics[width=\unitlength,page=4]{figure_states.pdf}}%
    \put(0.5554773,0.03612661){\color[rgb]{0.14901961,0.14901961,0.14901961}\makebox(0,0)[t]{\lineheight{1.25}\smash{\begin{tabular}[t]{c}0\end{tabular}}}}%
    \put(0.59793509,0.03612661){\color[rgb]{0.14901961,0.14901961,0.14901961}\makebox(0,0)[t]{\lineheight{1.25}\smash{\begin{tabular}[t]{c}0.5\end{tabular}}}}%
    \put(0.64039295,0.03612661){\color[rgb]{0.14901961,0.14901961,0.14901961}\makebox(0,0)[t]{\lineheight{1.25}\smash{\begin{tabular}[t]{c}1\end{tabular}}}}%
    \put(0.6828508,0.03612661){\color[rgb]{0.14901961,0.14901961,0.14901961}\makebox(0,0)[t]{\lineheight{1.25}\smash{\begin{tabular}[t]{c}1.5\end{tabular}}}}%
    \put(0.7253086,0.03612661){\color[rgb]{0.14901961,0.14901961,0.14901961}\makebox(0,0)[t]{\lineheight{1.25}\smash{\begin{tabular}[t]{c}2\end{tabular}}}}%
    \put(0.76776646,0.03612661){\color[rgb]{0.14901961,0.14901961,0.14901961}\makebox(0,0)[t]{\lineheight{1.25}\smash{\begin{tabular}[t]{c}2.5\end{tabular}}}}%
    \put(0.81022431,0.03612661){\color[rgb]{0.14901961,0.14901961,0.14901961}\makebox(0,0)[t]{\lineheight{1.25}\smash{\begin{tabular}[t]{c}3\end{tabular}}}}%
    \put(0.85268211,0.03612661){\color[rgb]{0.14901961,0.14901961,0.14901961}\makebox(0,0)[t]{\lineheight{1.25}\smash{\begin{tabular}[t]{c}3.5\end{tabular}}}}%
    \put(0.89513996,0.03612661){\color[rgb]{0.14901961,0.14901961,0.14901961}\makebox(0,0)[t]{\lineheight{1.25}\smash{\begin{tabular}[t]{c}4\end{tabular}}}}%
    \put(0.93759776,0.03612661){\color[rgb]{0.14901961,0.14901961,0.14901961}\makebox(0,0)[t]{\lineheight{1.25}\smash{\begin{tabular}[t]{c}4.5\end{tabular}}}}%
    \put(0.98005562,0.03612661){\color[rgb]{0.14901961,0.14901961,0.14901961}\makebox(0,0)[t]{\lineheight{1.25}\smash{\begin{tabular}[t]{c}5\end{tabular}}}}%
    \put(0,0){\includegraphics[width=\unitlength,page=5]{figure_states.pdf}}%
    \put(0.54698574,0.05863907){\color[rgb]{0.14901961,0.14901961,0.14901961}\makebox(0,0)[rt]{\lineheight{1.25}\smash{\begin{tabular}[t]{r}-2\end{tabular}}}}%
    \put(0.54698574,0.0881864){\color[rgb]{0.14901961,0.14901961,0.14901961}\makebox(0,0)[rt]{\lineheight{1.25}\smash{\begin{tabular}[t]{r}-1\end{tabular}}}}%
    \put(0.54698574,0.11773372){\color[rgb]{0.14901961,0.14901961,0.14901961}\makebox(0,0)[rt]{\lineheight{1.25}\smash{\begin{tabular}[t]{r}0\end{tabular}}}}%
    \put(0.54698574,0.14728105){\color[rgb]{0.14901961,0.14901961,0.14901961}\makebox(0,0)[rt]{\lineheight{1.25}\smash{\begin{tabular}[t]{r}1\end{tabular}}}}%
    \put(0,0){\includegraphics[width=\unitlength,page=6]{figure_states.pdf}}%
    \put(0.50979569,0.23330683){\color[rgb]{0.14901961,0.14901961,0.14901961}\rotatebox{90}{\makebox(0,0)[t]{\lineheight{1.25}\smash{\begin{tabular}[t]{c}$\omega_y$ (rad/s)\end{tabular}}}}}%
    \put(0.50979569,0.35642083){\color[rgb]{0.14901961,0.14901961,0.14901961}\rotatebox{90}{\makebox(0,0)[t]{\lineheight{1.25}\smash{\begin{tabular}[t]{c}$\omega_x$ (rad/s)\end{tabular}}}}}%
    \put(0.50686082,0.11019275){\color[rgb]{0.14901961,0.14901961,0.14901961}\rotatebox{90}{\makebox(0,0)[t]{\lineheight{1.25}\smash{\begin{tabular}[t]{c}$\omega_z$ (rad/s)\end{tabular}}}}}%
    \put(0,0){\includegraphics[width=\unitlength,page=7]{figure_states.pdf}}%
    \put(0.54642601,0.58620721){\color[rgb]{0.14901961,0.14901961,0.14901961}\makebox(0,0)[rt]{\lineheight{1.25}\smash{\begin{tabular}[t]{r}-5\end{tabular}}}}%
    \put(0.54642601,0.62155287){\color[rgb]{0.14901961,0.14901961,0.14901961}\makebox(0,0)[rt]{\lineheight{1.25}\smash{\begin{tabular}[t]{r}0\end{tabular}}}}%
    \put(0.54642601,0.65689853){\color[rgb]{0.14901961,0.14901961,0.14901961}\makebox(0,0)[rt]{\lineheight{1.25}\smash{\begin{tabular}[t]{r}5\end{tabular}}}}%
    \put(0,0){\includegraphics[width=\unitlength,page=8]{figure_states.pdf}}%
    \put(0.54642601,0.69741934){\color[rgb]{0.14901961,0.14901961,0.14901961}\makebox(0,0)[rt]{\lineheight{1.25}\smash{\begin{tabular}[t]{r}-5\end{tabular}}}}%
    \put(0.54642601,0.74155984){\color[rgb]{0.14901961,0.14901961,0.14901961}\makebox(0,0)[rt]{\lineheight{1.25}\smash{\begin{tabular}[t]{r}0\end{tabular}}}}%
    \put(0.54642601,0.78569935){\color[rgb]{0.14901961,0.14901961,0.14901961}\makebox(0,0)[rt]{\lineheight{1.25}\smash{\begin{tabular}[t]{r}5\end{tabular}}}}%
    \put(0,0){\includegraphics[width=\unitlength,page=9]{figure_states.pdf}}%
    \put(0.54642601,0.4636127){\color[rgb]{0.14901961,0.14901961,0.14901961}\makebox(0,0)[rt]{\lineheight{1.25}\smash{\begin{tabular}[t]{r}-5\end{tabular}}}}%
    \put(0.54642601,0.50172303){\color[rgb]{0.14901961,0.14901961,0.14901961}\makebox(0,0)[rt]{\lineheight{1.25}\smash{\begin{tabular}[t]{r}0\end{tabular}}}}%
    \put(0.54642601,0.53983431){\color[rgb]{0.14901961,0.14901961,0.14901961}\makebox(0,0)[rt]{\lineheight{1.25}\smash{\begin{tabular}[t]{r}5\end{tabular}}}}%
    \put(0,0){\includegraphics[width=\unitlength,page=10]{figure_states.pdf}}%
    \put(0.50879739,0.62162462){\color[rgb]{0.14901961,0.14901961,0.14901961}\rotatebox{90}{\makebox(0,0)[t]{\lineheight{1.25}\smash{\begin{tabular}[t]{c}$v_y$ (m/s)\end{tabular}}}}}%
    \put(0.50879739,0.74473861){\color[rgb]{0.14901961,0.14901961,0.14901961}\rotatebox{90}{\makebox(0,0)[t]{\lineheight{1.25}\smash{\begin{tabular}[t]{c}$v_x$ (m/s)\end{tabular}}}}}%
    \put(0.50879739,0.49851062){\color[rgb]{0.14901961,0.14901961,0.14901961}\rotatebox{90}{\makebox(0,0)[t]{\lineheight{1.25}\smash{\begin{tabular}[t]{c}$v_z$ (m/s)\end{tabular}}}}}%
    \put(0,0){\includegraphics[width=\unitlength,page=11]{figure_states.pdf}}%
    \put(0.04675811,0.18173906){\color[rgb]{0.14901961,0.14901961,0.14901961}\makebox(0,0)[rt]{\lineheight{1.25}\smash{\begin{tabular}[t]{r}-50\end{tabular}}}}%
    \put(0.04675811,0.21915672){\color[rgb]{0.14901961,0.14901961,0.14901961}\makebox(0,0)[rt]{\lineheight{1.25}\smash{\begin{tabular}[t]{r}0\end{tabular}}}}%
    \put(0.04675811,0.25657439){\color[rgb]{0.14901961,0.14901961,0.14901961}\makebox(0,0)[rt]{\lineheight{1.25}\smash{\begin{tabular}[t]{r}50\end{tabular}}}}%
    \put(0,0){\includegraphics[width=\unitlength,page=12]{figure_states.pdf}}%
    \put(0.04675811,0.30398039){\color[rgb]{0.14901961,0.14901961,0.14901961}\makebox(0,0)[rt]{\lineheight{1.25}\smash{\begin{tabular}[t]{r}-200\end{tabular}}}}%
    \put(0.04675811,0.34830186){\color[rgb]{0.14901961,0.14901961,0.14901961}\makebox(0,0)[rt]{\lineheight{1.25}\smash{\begin{tabular}[t]{r}0\end{tabular}}}}%
    \put(0.04675811,0.39262238){\color[rgb]{0.14901961,0.14901961,0.14901961}\makebox(0,0)[rt]{\lineheight{1.25}\smash{\begin{tabular}[t]{r}200\end{tabular}}}}%
    \put(0,0){\includegraphics[width=\unitlength,page=13]{figure_states.pdf}}%
    \put(0.26753881,0.00258557){\color[rgb]{0.14901961,0.14901961,0.14901961}\makebox(0,0)[t]{\lineheight{1.25}\smash{\begin{tabular}[t]{c}Time (s)\end{tabular}}}}%
    \put(0,0){\includegraphics[width=\unitlength,page=14]{figure_states.pdf}}%
    \put(0.05524967,0.03524093){\color[rgb]{0.14901961,0.14901961,0.14901961}\makebox(0,0)[t]{\lineheight{1.25}\smash{\begin{tabular}[t]{c}0\end{tabular}}}}%
    \put(0.09770751,0.03524093){\color[rgb]{0.14901961,0.14901961,0.14901961}\makebox(0,0)[t]{\lineheight{1.25}\smash{\begin{tabular}[t]{c}0.5\end{tabular}}}}%
    \put(0.14016533,0.03524093){\color[rgb]{0.14901961,0.14901961,0.14901961}\makebox(0,0)[t]{\lineheight{1.25}\smash{\begin{tabular}[t]{c}1\end{tabular}}}}%
    \put(0.18262316,0.03524093){\color[rgb]{0.14901961,0.14901961,0.14901961}\makebox(0,0)[t]{\lineheight{1.25}\smash{\begin{tabular}[t]{c}1.5\end{tabular}}}}%
    \put(0.22508098,0.03524093){\color[rgb]{0.14901961,0.14901961,0.14901961}\makebox(0,0)[t]{\lineheight{1.25}\smash{\begin{tabular}[t]{c}2\end{tabular}}}}%
    \put(0.26753881,0.03524093){\color[rgb]{0.14901961,0.14901961,0.14901961}\makebox(0,0)[t]{\lineheight{1.25}\smash{\begin{tabular}[t]{c}2.5\end{tabular}}}}%
    \put(0.30999667,0.03524093){\color[rgb]{0.14901961,0.14901961,0.14901961}\makebox(0,0)[t]{\lineheight{1.25}\smash{\begin{tabular}[t]{c}3\end{tabular}}}}%
    \put(0.35245449,0.03524093){\color[rgb]{0.14901961,0.14901961,0.14901961}\makebox(0,0)[t]{\lineheight{1.25}\smash{\begin{tabular}[t]{c}3.5\end{tabular}}}}%
    \put(0.39491232,0.03524093){\color[rgb]{0.14901961,0.14901961,0.14901961}\makebox(0,0)[t]{\lineheight{1.25}\smash{\begin{tabular}[t]{c}4\end{tabular}}}}%
    \put(0.43737015,0.03524093){\color[rgb]{0.14901961,0.14901961,0.14901961}\makebox(0,0)[t]{\lineheight{1.25}\smash{\begin{tabular}[t]{c}4.5\end{tabular}}}}%
    \put(0.47933924,0.03577193){\color[rgb]{0.14901961,0.14901961,0.14901961}\makebox(0,0)[t]{\lineheight{1.25}\smash{\begin{tabular}[t]{c}5\end{tabular}}}}%
    \put(0,0){\includegraphics[width=\unitlength,page=15]{figure_states.pdf}}%
    \put(0.04675811,0.05775339){\color[rgb]{0.14901961,0.14901961,0.14901961}\makebox(0,0)[rt]{\lineheight{1.25}\smash{\begin{tabular}[t]{r}-200\end{tabular}}}}%
    \put(0.04675811,0.10207384){\color[rgb]{0.14901961,0.14901961,0.14901961}\makebox(0,0)[rt]{\lineheight{1.25}\smash{\begin{tabular}[t]{r}0\end{tabular}}}}%
    \put(0.04675811,0.14639537){\color[rgb]{0.14901961,0.14901961,0.14901961}\makebox(0,0)[rt]{\lineheight{1.25}\smash{\begin{tabular}[t]{r}200\end{tabular}}}}%
    \put(0,0){\includegraphics[width=\unitlength,page=16]{figure_states.pdf}}%
    \put(0.01036926,0.22987034){\color[rgb]{0.14901961,0.14901961,0.14901961}\rotatebox{90}{\makebox(0,0)[t]{\lineheight{1.25}\smash{\begin{tabular}[t]{c}$\theta$ (deg)\end{tabular}}}}}%
    \put(0.01134755,0.35298433){\color[rgb]{0.14901961,0.14901961,0.14901961}\rotatebox{90}{\makebox(0,0)[t]{\lineheight{1.25}\smash{\begin{tabular}[t]{c}$\phi$ (deg)\end{tabular}}}}}%
    \put(0.00939097,0.10675631){\color[rgb]{0.14901961,0.14901961,0.14901961}\rotatebox{90}{\makebox(0,0)[t]{\lineheight{1.25}\smash{\begin{tabular}[t]{c}$\psi$ (deg)\end{tabular}}}}}%
    \put(0,0){\includegraphics[width=\unitlength,page=17]{figure_states.pdf}}%
    \put(0.04835239,0.57138932){\color[rgb]{0.14901961,0.14901961,0.14901961}\makebox(0,0)[rt]{\lineheight{1.25}\smash{\begin{tabular}[t]{r}-5\end{tabular}}}}%
    \put(0.04835239,0.61568634){\color[rgb]{0.14901961,0.14901961,0.14901961}\makebox(0,0)[rt]{\lineheight{1.25}\smash{\begin{tabular}[t]{r}0\end{tabular}}}}%
    \put(0.04835239,0.65998434){\color[rgb]{0.14901961,0.14901961,0.14901961}\makebox(0,0)[rt]{\lineheight{1.25}\smash{\begin{tabular}[t]{r}5\end{tabular}}}}%
    \put(0,0){\includegraphics[width=\unitlength,page=18]{figure_states.pdf}}%
    \put(0.04835239,0.69450236){\color[rgb]{0.14901961,0.14901961,0.14901961}\makebox(0,0)[rt]{\lineheight{1.25}\smash{\begin{tabular}[t]{r}-5\end{tabular}}}}%
    \put(0.04835239,0.73882382){\color[rgb]{0.14901961,0.14901961,0.14901961}\makebox(0,0)[rt]{\lineheight{1.25}\smash{\begin{tabular}[t]{r}0\end{tabular}}}}%
    \put(0.04835239,0.78314432){\color[rgb]{0.14901961,0.14901961,0.14901961}\makebox(0,0)[rt]{\lineheight{1.25}\smash{\begin{tabular}[t]{r}5\end{tabular}}}}%
    \put(0,0){\includegraphics[width=\unitlength,page=19]{figure_states.pdf}}%
    \put(0.04835239,0.44827531){\color[rgb]{0.14901961,0.14901961,0.14901961}\makebox(0,0)[rt]{\lineheight{1.25}\smash{\begin{tabular}[t]{r}0\end{tabular}}}}%
    \put(0.04835239,0.49259579){\color[rgb]{0.14901961,0.14901961,0.14901961}\makebox(0,0)[rt]{\lineheight{1.25}\smash{\begin{tabular}[t]{r}5\end{tabular}}}}%
    \put(0.04835239,0.53691729){\color[rgb]{0.14901961,0.14901961,0.14901961}\makebox(0,0)[rt]{\lineheight{1.25}\smash{\begin{tabular}[t]{r}10\end{tabular}}}}%
    \put(0,0){\includegraphics[width=\unitlength,page=20]{figure_states.pdf}}%
    \put(0.00975325,0.62596792){\color[rgb]{0.14901961,0.14901961,0.14901961}\rotatebox{90}{\makebox(0,0)[t]{\lineheight{1.25}\smash{\begin{tabular}[t]{c}$\xi_y$ (m)\end{tabular}}}}}%
    \put(0.00975325,0.74908191){\color[rgb]{0.14901961,0.14901961,0.14901961}\rotatebox{90}{\makebox(0,0)[t]{\lineheight{1.25}\smash{\begin{tabular}[t]{c}$\xi_x$ (m)\end{tabular}}}}}%
    \put(0.00975325,0.50285388){\color[rgb]{0.14901961,0.14901961,0.14901961}\rotatebox{90}{\makebox(0,0)[t]{\lineheight{1.25}\smash{\begin{tabular}[t]{c}$\xi_z$ (m)\end{tabular}}}}}%
    \put(0,0){\includegraphics[width=\unitlength,page=21]{figure_states.pdf}}%
    \put(0.88018705,0.19236066){\color[rgb]{0,0,0}\makebox(0,0)[lt]{\lineheight{1.25}\smash{\begin{tabular}[t]{l}$x_{04}$FMNMPC\end{tabular}}}}%
    \put(0,0){\includegraphics[width=\unitlength,page=22]{figure_states.pdf}}%
    \put(0.88018705,0.21143734){\color[rgb]{0,0,0}\makebox(0,0)[lt]{\lineheight{1.25}\smash{\begin{tabular}[t]{l}$x_{04}$NMPC\end{tabular}}}}%
    \put(0.88024957,0.23257706){\color[rgb]{0,0,0}\makebox(0,0)[lt]{\lineheight{1.25}\smash{\begin{tabular}[t]{l}$x_{03}$FMNMPC\end{tabular}}}}%
    \put(0.87943417,0.27146669){\color[rgb]{0,0,0}\makebox(0,0)[lt]{\lineheight{1.25}\smash{\begin{tabular}[t]{l}$x_{02}$FMNMPC\end{tabular}}}}%
    \put(0.87988056,0.30684454){\color[rgb]{0,0,0}\makebox(0,0)[lt]{\lineheight{1.25}\smash{\begin{tabular}[t]{l}$x_{01}$FMNMPC\end{tabular}}}}%
    \put(0.87966244,0.25188547){\color[rgb]{0,0,0}\makebox(0,0)[lt]{\lineheight{1.25}\smash{\begin{tabular}[t]{l}$x_{03}$NMPC\end{tabular}}}}%
    \put(0.87982031,0.28990137){\color[rgb]{0,0,0}\makebox(0,0)[lt]{\lineheight{1.25}\smash{\begin{tabular}[t]{l}$x_{02}$NMPC\end{tabular}}}}%
    \put(0.87944873,0.3221195){\color[rgb]{0,0,0}\makebox(0,0)[lt]{\lineheight{1.25}\smash{\begin{tabular}[t]{l}$x_{01}$NMPC\end{tabular}}}}%
    \put(0,0){\includegraphics[width=\unitlength,page=23]{figure_states.pdf}}%
  \end{picture}%
\endgroup%

%% file: Figuras/figure_forces.pdf_tex
\tiny
\begingroup%
  \makeatletter%
  \providecommand\color[2][]{%
    \errmessage{(Inkscape) Color is used for the text in Inkscape, but the package 'color.sty' is not loaded}%
    \renewcommand\color[2][]{}%
  }%
  \providecommand\transparent[1]{%
    \errmessage{(Inkscape) Transparency is used (non-zero) for the text in Inkscape, but the package 'transparent.sty' is not loaded}%
    \renewcommand\transparent[1]{}%
  }%
  \providecommand\rotatebox[2]{#2}%
  \newcommand*\fsize{\dimexpr\f@size pt\relax}%
  \newcommand*\lineheight[1]{\fontsize{\fsize}{#1\fsize}\selectfont}%
  \ifx\svgwidth\undefined%
    \setlength{\unitlength}{364.60901642bp}%
    \ifx\svgscale\undefined%
      \relax%
    \else%
      \setlength{\unitlength}{\unitlength * \real{\svgscale}}%
    \fi%
  \else%
    \setlength{\unitlength}{\svgwidth}%
  \fi%
  \global\let\svgwidth\undefined%
  \global\let\svgscale\undefined%
  \makeatother%
  \begin{picture}(1,0.77311196)%
    \lineheight{1}%
    \setlength\tabcolsep{0pt}%
    \put(0,0){\includegraphics[width=\unitlength,page=1]{figure_forces.pdf}}%
    \put(0.07504045,0.46292219){\color[rgb]{0.14901961,0.14901961,0.14901961}\makebox(0,0)[rt]{\lineheight{1.25}\smash{\begin{tabular}[t]{r}0\end{tabular}}}}%
    \put(0.07504045,0.50780794){\color[rgb]{0.14901961,0.14901961,0.14901961}\makebox(0,0)[rt]{\lineheight{1.25}\smash{\begin{tabular}[t]{r}5\end{tabular}}}}%
    \put(0.07504045,0.55269569){\color[rgb]{0.14901961,0.14901961,0.14901961}\makebox(0,0)[rt]{\lineheight{1.25}\smash{\begin{tabular}[t]{r}10\end{tabular}}}}%
    \put(0,0){\includegraphics[width=\unitlength,page=2]{figure_forces.pdf}}%
    \put(0.07504045,0.65219891){\color[rgb]{0.14901961,0.14901961,0.14901961}\makebox(0,0)[rt]{\lineheight{1.25}\smash{\begin{tabular}[t]{r}0\end{tabular}}}}%
    \put(0.07504045,0.69708466){\color[rgb]{0.14901961,0.14901961,0.14901961}\makebox(0,0)[rt]{\lineheight{1.25}\smash{\begin{tabular}[t]{r}5\end{tabular}}}}%
    \put(0.07504045,0.74197247){\color[rgb]{0.14901961,0.14901961,0.14901961}\makebox(0,0)[rt]{\lineheight{1.25}\smash{\begin{tabular}[t]{r}10\end{tabular}}}}%
    \put(0,0){\includegraphics[width=\unitlength,page=3]{figure_forces.pdf}}%
    \put(0.07504045,0.27364548){\color[rgb]{0.14901961,0.14901961,0.14901961}\makebox(0,0)[rt]{\lineheight{1.25}\smash{\begin{tabular}[t]{r}0\end{tabular}}}}%
    \put(0.07504045,0.3185312){\color[rgb]{0.14901961,0.14901961,0.14901961}\makebox(0,0)[rt]{\lineheight{1.25}\smash{\begin{tabular}[t]{r}5\end{tabular}}}}%
    \put(0.07504045,0.36341905){\color[rgb]{0.14901961,0.14901961,0.14901961}\makebox(0,0)[rt]{\lineheight{1.25}\smash{\begin{tabular}[t]{r}10\end{tabular}}}}%
    \put(0,0){\includegraphics[width=\unitlength,page=4]{figure_forces.pdf}}%
    \put(0.53926372,0.00511669){\color[rgb]{0.14901961,0.14901961,0.14901961}\makebox(0,0)[t]{\lineheight{1.25}\smash{\begin{tabular}[t]{c}Time (s)\end{tabular}}}}%
    \put(0,0){\includegraphics[width=\unitlength,page=5]{figure_forces.pdf}}%
    \put(0.09289519,0.05089312){\color[rgb]{0.14901961,0.14901961,0.14901961}\makebox(0,0)[t]{\lineheight{1.25}\smash{\begin{tabular}[t]{c}0\end{tabular}}}}%
    \put(0.18216889,0.05089312){\color[rgb]{0.14901961,0.14901961,0.14901961}\makebox(0,0)[t]{\lineheight{1.25}\smash{\begin{tabular}[t]{c}0.5\end{tabular}}}}%
    \put(0.27144262,0.05089312){\color[rgb]{0.14901961,0.14901961,0.14901961}\makebox(0,0)[t]{\lineheight{1.25}\smash{\begin{tabular}[t]{c}1\end{tabular}}}}%
    \put(0.36071632,0.05089312){\color[rgb]{0.14901961,0.14901961,0.14901961}\makebox(0,0)[t]{\lineheight{1.25}\smash{\begin{tabular}[t]{c}1.5\end{tabular}}}}%
    \put(0.44999002,0.05089312){\color[rgb]{0.14901961,0.14901961,0.14901961}\makebox(0,0)[t]{\lineheight{1.25}\smash{\begin{tabular}[t]{c}2\end{tabular}}}}%
    \put(0.53926372,0.05089312){\color[rgb]{0.14901961,0.14901961,0.14901961}\makebox(0,0)[t]{\lineheight{1.25}\smash{\begin{tabular}[t]{c}2.5\end{tabular}}}}%
    \put(0.62853748,0.05089312){\color[rgb]{0.14901961,0.14901961,0.14901961}\makebox(0,0)[t]{\lineheight{1.25}\smash{\begin{tabular}[t]{c}3\end{tabular}}}}%
    \put(0.71781118,0.05089312){\color[rgb]{0.14901961,0.14901961,0.14901961}\makebox(0,0)[t]{\lineheight{1.25}\smash{\begin{tabular}[t]{c}3.5\end{tabular}}}}%
    \put(0.80708488,0.05089312){\color[rgb]{0.14901961,0.14901961,0.14901961}\makebox(0,0)[t]{\lineheight{1.25}\smash{\begin{tabular}[t]{c}4\end{tabular}}}}%
    \put(0.89635858,0.05089312){\color[rgb]{0.14901961,0.14901961,0.14901961}\makebox(0,0)[t]{\lineheight{1.25}\smash{\begin{tabular}[t]{c}4.5\end{tabular}}}}%
    \put(0.98563228,0.05089312){\color[rgb]{0.14901961,0.14901961,0.14901961}\makebox(0,0)[t]{\lineheight{1.25}\smash{\begin{tabular}[t]{c}5\end{tabular}}}}%
    \put(0,0){\includegraphics[width=\unitlength,page=6]{figure_forces.pdf}}%
    \put(0.07504045,0.08436667){\color[rgb]{0.14901961,0.14901961,0.14901961}\makebox(0,0)[rt]{\lineheight{1.25}\smash{\begin{tabular}[t]{r}0\end{tabular}}}}%
    \put(0.07504045,0.12925446){\color[rgb]{0.14901961,0.14901961,0.14901961}\makebox(0,0)[rt]{\lineheight{1.25}\smash{\begin{tabular}[t]{r}5\end{tabular}}}}%
    \put(0.07504045,0.17414225){\color[rgb]{0.14901961,0.14901961,0.14901961}\makebox(0,0)[rt]{\lineheight{1.25}\smash{\begin{tabular}[t]{r}10\end{tabular}}}}%
    \put(0,0){\includegraphics[width=\unitlength,page=7]{figure_forces.pdf}}%
    \put(0.01858424,0.52501462){\color[rgb]{0.14901961,0.14901961,0.14901961}\rotatebox{90}{\makebox(0,0)[t]{\lineheight{1.25}\smash{\begin{tabular}[t]{c}$f_2$ (N)\end{tabular}}}}}%
    \put(0.01858424,0.71429135){\color[rgb]{0.14901961,0.14901961,0.14901961}\rotatebox{90}{\makebox(0,0)[t]{\lineheight{1.25}\smash{\begin{tabular}[t]{c}$f_1$ (N)\end{tabular}}}}}%
    \put(0.01858424,0.33573797){\color[rgb]{0.14901961,0.14901961,0.14901961}\rotatebox{90}{\makebox(0,0)[t]{\lineheight{1.25}\smash{\begin{tabular}[t]{c}$f_3$ (N)\end{tabular}}}}}%
    \put(0.01858424,0.1464611){\color[rgb]{0.14901961,0.14901961,0.14901961}\rotatebox{90}{\makebox(0,0)[t]{\lineheight{1.25}\smash{\begin{tabular}[t]{c}$f_4$ (N)\end{tabular}}}}}%
    \put(0,0){\includegraphics[width=\unitlength,page=8]{figure_forces.pdf}}%
    \put(0.78734089,0.40513371){\color[rgb]{0,0,0}\makebox(0,0)[lt]{\lineheight{1.25}\smash{\begin{tabular}[t]{l}$x_{04}$FMNMPC\end{tabular}}}}%
    \put(0,0){\includegraphics[width=\unitlength,page=9]{figure_forces.pdf}}%
    \put(0.78734089,0.43290319){\color[rgb]{0,0,0}\makebox(0,0)[lt]{\lineheight{1.25}\smash{\begin{tabular}[t]{l}$x_{04}$NMPC\end{tabular}}}}%
    \put(0.78747234,0.46089648){\color[rgb]{0,0,0}\makebox(0,0)[lt]{\lineheight{1.25}\smash{\begin{tabular}[t]{l}$x_{03}$FMNMPC\end{tabular}}}}%
    \put(0.78575784,0.51798359){\color[rgb]{0,0,0}\makebox(0,0)[lt]{\lineheight{1.25}\smash{\begin{tabular}[t]{l}$x_{02}$FMNMPC\end{tabular}}}}%
    \put(0.78669644,0.56768663){\color[rgb]{0,0,0}\makebox(0,0)[lt]{\lineheight{1.25}\smash{\begin{tabular}[t]{l}$x_{01}$FMNMPC\end{tabular}}}}%
    \put(0.78623769,0.48915327){\color[rgb]{0,0,0}\makebox(0,0)[lt]{\lineheight{1.25}\smash{\begin{tabular}[t]{l}$x_{03}$NMPC\end{tabular}}}}%
    \put(0.78656976,0.54440312){\color[rgb]{0,0,0}\makebox(0,0)[lt]{\lineheight{1.25}\smash{\begin{tabular}[t]{l}$x_{02}$NMPC\end{tabular}}}}%
    \put(0.7857886,0.59569044){\color[rgb]{0,0,0}\makebox(0,0)[lt]{\lineheight{1.25}\smash{\begin{tabular}[t]{l}$x_{01}$NMPC\end{tabular}}}}%
  \end{picture}%
\endgroup%

%% file: Figuras/figure910_small.pdf_tex
\begingroup%
  \makeatletter%
  \providecommand\color[2][]{%
    \errmessage{(Inkscape) Color is used for the text in Inkscape, but the package 'color.sty' is not loaded}%
    \renewcommand\color[2][]{}%
  }%
  \providecommand\transparent[1]{%
    \errmessage{(Inkscape) Transparency is used (non-zero) for the text in Inkscape, but the package 'transparent.sty' is not loaded}%
    \renewcommand\transparent[1]{}%
  }%
  \providecommand\rotatebox[2]{#2}%
  \newcommand*\fsize{\dimexpr\f@size pt\relax}%
  \newcommand*\lineheight[1]{\fontsize{\fsize}{#1\fsize}\selectfont}%
  \ifx\svgwidth\undefined%
    \setlength{\unitlength}{797.61209106bp}%
    \ifx\svgscale\undefined%
      \relax%
    \else%
      \setlength{\unitlength}{\unitlength * \real{\svgscale}}%
    \fi%
  \else%
    \setlength{\unitlength}{\svgwidth}%
  \fi%
  \global\let\svgwidth\undefined%
  \global\let\svgscale\undefined%
  \makeatother%
  \begin{picture}(1,0.39248221)%
    \lineheight{1}%
    \setlength\tabcolsep{0pt}%
    \put(0,0){\includegraphics[width=\unitlength,page=1]{figure910_small.pdf}}%
    \put(0.08772174,0.23641007){\color[rgb]{0.14901961,0.14901961,0.14901961}\makebox(0,0)[rt]{\lineheight{1.25}\smash{\begin{tabular}[t]{r}0\end{tabular}}}}%
    \put(0.08772174,0.26755867){\color[rgb]{0.14901961,0.14901961,0.14901961}\makebox(0,0)[rt]{\lineheight{1.25}\smash{\begin{tabular}[t]{r}5000\end{tabular}}}}%
    \put(0.08772174,0.29870728){\color[rgb]{0.14901961,0.14901961,0.14901961}\makebox(0,0)[rt]{\lineheight{1.25}\smash{\begin{tabular}[t]{r}10000\end{tabular}}}}%
    \put(0,0){\includegraphics[width=\unitlength,page=2]{figure910_small.pdf}}%
    \put(0.08772174,0.32293334){\color[rgb]{0.14901961,0.14901961,0.14901961}\makebox(0,0)[rt]{\lineheight{1.25}\smash{\begin{tabular}[t]{r}0\end{tabular}}}}%
    \put(0.08772174,0.35408194){\color[rgb]{0.14901961,0.14901961,0.14901961}\makebox(0,0)[rt]{\lineheight{1.25}\smash{\begin{tabular}[t]{r}5000\end{tabular}}}}%
    \put(0.08772174,0.38523054){\color[rgb]{0.14901961,0.14901961,0.14901961}\makebox(0,0)[rt]{\lineheight{1.25}\smash{\begin{tabular}[t]{r}10000\end{tabular}}}}%
    \put(0,0){\includegraphics[width=\unitlength,page=3]{figure910_small.pdf}}%
    \put(0.08772174,0.14988681){\color[rgb]{0.14901961,0.14901961,0.14901961}\makebox(0,0)[rt]{\lineheight{1.25}\smash{\begin{tabular}[t]{r}0\end{tabular}}}}%
    \put(0.08772174,0.18992507){\color[rgb]{0.14901961,0.14901961,0.14901961}\makebox(0,0)[rt]{\lineheight{1.25}\smash{\begin{tabular}[t]{r}5000\end{tabular}}}}%
    \put(0,0){\includegraphics[width=\unitlength,page=4]{figure910_small.pdf}}%
    \put(0.09588361,0.0497056){\color[rgb]{0.14901961,0.14901961,0.14901961}\makebox(0,0)[t]{\lineheight{1.25}\smash{\begin{tabular}[t]{c}0\end{tabular}}}}%
    \put(0.13669291,0.0497056){\color[rgb]{0.14901961,0.14901961,0.14901961}\makebox(0,0)[t]{\lineheight{1.25}\smash{\begin{tabular}[t]{c}0.5\end{tabular}}}}%
    \put(0.17750223,0.0497056){\color[rgb]{0.14901961,0.14901961,0.14901961}\makebox(0,0)[t]{\lineheight{1.25}\smash{\begin{tabular}[t]{c}1\end{tabular}}}}%
    \put(0.21831154,0.0497056){\color[rgb]{0.14901961,0.14901961,0.14901961}\makebox(0,0)[t]{\lineheight{1.25}\smash{\begin{tabular}[t]{c}1.5\end{tabular}}}}%
    \put(0.25912084,0.0497056){\color[rgb]{0.14901961,0.14901961,0.14901961}\makebox(0,0)[t]{\lineheight{1.25}\smash{\begin{tabular}[t]{c}2\end{tabular}}}}%
    \put(0.29993015,0.0497056){\color[rgb]{0.14901961,0.14901961,0.14901961}\makebox(0,0)[t]{\lineheight{1.25}\smash{\begin{tabular}[t]{c}2.5\end{tabular}}}}%
    \put(0.34073948,0.0497056){\color[rgb]{0.14901961,0.14901961,0.14901961}\makebox(0,0)[t]{\lineheight{1.25}\smash{\begin{tabular}[t]{c}3\end{tabular}}}}%
    \put(0.38154879,0.0497056){\color[rgb]{0.14901961,0.14901961,0.14901961}\makebox(0,0)[t]{\lineheight{1.25}\smash{\begin{tabular}[t]{c}3.5\end{tabular}}}}%
    \put(0.42235809,0.0497056){\color[rgb]{0.14901961,0.14901961,0.14901961}\makebox(0,0)[t]{\lineheight{1.25}\smash{\begin{tabular}[t]{c}4\end{tabular}}}}%
    \put(0.4631674,0.0497056){\color[rgb]{0.14901961,0.14901961,0.14901961}\makebox(0,0)[t]{\lineheight{1.25}\smash{\begin{tabular}[t]{c}4.5\end{tabular}}}}%
    \put(0.50397668,0.0497056){\color[rgb]{0.14901961,0.14901961,0.14901961}\makebox(0,0)[t]{\lineheight{1.25}\smash{\begin{tabular}[t]{c}5\end{tabular}}}}%
    \put(0,0){\includegraphics[width=\unitlength,page=5]{figure910_small.pdf}}%
    \put(0.08772174,0.06336355){\color[rgb]{0.14901961,0.14901961,0.14901961}\makebox(0,0)[rt]{\lineheight{1.25}\smash{\begin{tabular}[t]{r}0\end{tabular}}}}%
    \put(0.08772174,0.10376288){\color[rgb]{0.14901961,0.14901961,0.14901961}\makebox(0,0)[rt]{\lineheight{1.25}\smash{\begin{tabular}[t]{r}5000\end{tabular}}}}%
    \put(0,0){\includegraphics[width=\unitlength,page=6]{figure910_small.pdf}}%
    \put(0.28377781,0.00367261){\makebox(0,0)[lt]{\lineheight{1.25}\smash{\begin{tabular}[t]{l}Time (s)\end{tabular}}}}%
    \put(0,0){\includegraphics[width=\unitlength,page=7]{figure910_small.pdf}}%
    \put(0.37565589,0.08833715){\color[rgb]{0,0,0}\makebox(0,0)[lt]{\lineheight{1.25}\smash{\begin{tabular}[t]{l}\tiny FMNMPC\end{tabular}}}}%
    \put(0,0){\includegraphics[width=\unitlength,page=8]{figure910_small.pdf}}%
    \put(0.37565589,0.10949405){\color[rgb]{0,0,0}\makebox(0,0)[lt]{\lineheight{1.25}\smash{\begin{tabular}[t]{l}\tiny NMPC\end{tabular}}}}%
    \put(0,0){\includegraphics[width=\unitlength,page=9]{figure910_small.pdf}}%
    \put(0.37565589,0.1748604){\color[rgb]{0,0,0}\makebox(0,0)[lt]{\lineheight{1.25}\smash{\begin{tabular}[t]{l}\tiny FMNMPC\end{tabular}}}}%
    \put(0.37565589,0.1960173){\color[rgb]{0,0,0}\makebox(0,0)[lt]{\lineheight{1.25}\smash{\begin{tabular}[t]{l}\tiny NMPC\end{tabular}}}}%
    \put(0,0){\includegraphics[width=\unitlength,page=10]{figure910_small.pdf}}%
    \put(0.37565589,0.25950306){\color[rgb]{0,0,0}\makebox(0,0)[lt]{\lineheight{1.25}\smash{\begin{tabular}[t]{l}\tiny FMNMPC\end{tabular}}}}%
    \put(0.37565589,0.28254059){\color[rgb]{0,0,0}\makebox(0,0)[lt]{\lineheight{1.25}\smash{\begin{tabular}[t]{l}\tiny NMPC\end{tabular}}}}%
    \put(0,0){\includegraphics[width=\unitlength,page=11]{figure910_small.pdf}}%
    \put(0.37565589,0.34602633){\color[rgb]{0,0,0}\makebox(0,0)[lt]{\lineheight{1.25}\smash{\begin{tabular}[t]{l}\tiny FMNMPC\end{tabular}}}}%
    \put(0.37565589,0.36906384){\color[rgb]{0,0,0}\makebox(0,0)[lt]{\lineheight{1.25}\smash{\begin{tabular}[t]{l}\tiny NMPC\end{tabular}}}}%
    \put(0,0){\includegraphics[width=\unitlength,page=12]{figure910_small.pdf}}%
    \put(0.00762278,0.16014165){\rotatebox{90}{\makebox(0,0)[lt]{\lineheight{1.25}\smash{\begin{tabular}[t]{l}$\|\bar{x}\|^{2}_{Q_{\bar{x}}}$\end{tabular}}}}}%
    \put(0,0){\includegraphics[width=\unitlength,page=13]{figure910_small.pdf}}%
    \put(0.58138795,0.23598219){\color[rgb]{0.14901961,0.14901961,0.14901961}\makebox(0,0)[rt]{\lineheight{1.25}\smash{\begin{tabular}[t]{r}0\end{tabular}}}}%
    \put(0.58138795,0.28072106){\color[rgb]{0.14901961,0.14901961,0.14901961}\makebox(0,0)[rt]{\lineheight{1.25}\smash{\begin{tabular}[t]{r}5\end{tabular}}}}%
    \put(0,0){\includegraphics[width=\unitlength,page=14]{figure910_small.pdf}}%
    \put(0.58138795,0.32250547){\color[rgb]{0.14901961,0.14901961,0.14901961}\makebox(0,0)[rt]{\lineheight{1.25}\smash{\begin{tabular}[t]{r}0\end{tabular}}}}%
    \put(0.58138795,0.366556){\color[rgb]{0.14901961,0.14901961,0.14901961}\makebox(0,0)[rt]{\lineheight{1.25}\smash{\begin{tabular}[t]{r}5\end{tabular}}}}%
    \put(0,0){\includegraphics[width=\unitlength,page=15]{figure910_small.pdf}}%
    \put(0.58954981,0.14144278){\color[rgb]{0.14901961,0.14901961,0.14901961}\makebox(0,0)[t]{\lineheight{1.25}\smash{\begin{tabular}[t]{c}0\end{tabular}}}}%
    \put(0,0){\includegraphics[width=\unitlength,page=16]{figure910_small.pdf}}%
    \put(0.58138795,0.14945892){\color[rgb]{0.14901961,0.14901961,0.14901961}\makebox(0,0)[rt]{\lineheight{1.25}\smash{\begin{tabular}[t]{r}0\end{tabular}}}}%
    \put(0.58138795,0.19349254){\color[rgb]{0.14901961,0.14901961,0.14901961}\makebox(0,0)[rt]{\lineheight{1.25}\smash{\begin{tabular}[t]{r}5\end{tabular}}}}%
    \put(0,0){\includegraphics[width=\unitlength,page=17]{figure910_small.pdf}}%
    \put(0.58954981,0.04927769){\color[rgb]{0.14901961,0.14901961,0.14901961}\makebox(0,0)[t]{\lineheight{1.25}\smash{\begin{tabular}[t]{c}0\end{tabular}}}}%
    \put(0.63035909,0.04927769){\color[rgb]{0.14901961,0.14901961,0.14901961}\makebox(0,0)[t]{\lineheight{1.25}\smash{\begin{tabular}[t]{c}0.5\end{tabular}}}}%
    \put(0.67116842,0.04927769){\color[rgb]{0.14901961,0.14901961,0.14901961}\makebox(0,0)[t]{\lineheight{1.25}\smash{\begin{tabular}[t]{c}1\end{tabular}}}}%
    \put(0.71197775,0.04927769){\color[rgb]{0.14901961,0.14901961,0.14901961}\makebox(0,0)[t]{\lineheight{1.25}\smash{\begin{tabular}[t]{c}1.5\end{tabular}}}}%
    \put(0.75278703,0.04927769){\color[rgb]{0.14901961,0.14901961,0.14901961}\makebox(0,0)[t]{\lineheight{1.25}\smash{\begin{tabular}[t]{c}2\end{tabular}}}}%
    \put(0.79359636,0.04927769){\color[rgb]{0.14901961,0.14901961,0.14901961}\makebox(0,0)[t]{\lineheight{1.25}\smash{\begin{tabular}[t]{c}2.5\end{tabular}}}}%
    \put(0.8344057,0.04927769){\color[rgb]{0.14901961,0.14901961,0.14901961}\makebox(0,0)[t]{\lineheight{1.25}\smash{\begin{tabular}[t]{c}3\end{tabular}}}}%
    \put(0.87521498,0.04927769){\color[rgb]{0.14901961,0.14901961,0.14901961}\makebox(0,0)[t]{\lineheight{1.25}\smash{\begin{tabular}[t]{c}3.5\end{tabular}}}}%
    \put(0.91602431,0.04927769){\color[rgb]{0.14901961,0.14901961,0.14901961}\makebox(0,0)[t]{\lineheight{1.25}\smash{\begin{tabular}[t]{c}4\end{tabular}}}}%
    \put(0.95683353,0.04927769){\color[rgb]{0.14901961,0.14901961,0.14901961}\makebox(0,0)[t]{\lineheight{1.25}\smash{\begin{tabular}[t]{c}4.5\end{tabular}}}}%
    \put(0.99764292,0.04927769){\color[rgb]{0.14901961,0.14901961,0.14901961}\makebox(0,0)[t]{\lineheight{1.25}\smash{\begin{tabular}[t]{c}5\end{tabular}}}}%
    \put(0,0){\includegraphics[width=\unitlength,page=18]{figure910_small.pdf}}%
    \put(0.58138795,0.06293567){\color[rgb]{0.14901961,0.14901961,0.14901961}\makebox(0,0)[rt]{\lineheight{1.25}\smash{\begin{tabular}[t]{r}0\end{tabular}}}}%
    \put(0.58138795,0.1069862){\color[rgb]{0.14901961,0.14901961,0.14901961}\makebox(0,0)[rt]{\lineheight{1.25}\smash{\begin{tabular}[t]{r}5\end{tabular}}}}%
    \put(0,0){\includegraphics[width=\unitlength,page=19]{figure910_small.pdf}}%
    \put(0.77492252,0.00158473){\makebox(0,0)[lt]{\lineheight{1.25}\smash{\begin{tabular}[t]{l}Time (s)\end{tabular}}}}%
    \put(0,0){\includegraphics[width=\unitlength,page=20]{figure910_small.pdf}}%
    \put(0.54279802,0.19680175){\rotatebox{90}{\makebox(0,0)[lt]{\lineheight{1.25}\smash{\begin{tabular}[t]{l}$\|\xi^{\mathcal{I}}\|$\end{tabular}}}}}%
  \end{picture}%
\endgroup%